\documentclass[pre,floatfix,twocolumn]{revtex4}

\usepackage{amsmath,amssymb,latexsym}
\usepackage{graphicx}
\usepackage{xcolor}
\usepackage{verbatim}

\newcommand{\Sex}{S_{\rm{ex}}}
\newcommand{\bfa}[1]{\mathbf{#1}}

\newcommand{\newpurple}[1]{{\color{purple}#1}}

\begin{document}

\title{Isomorphs in sheared binary Lennard-Jones glass: Transient response}

\date{\today}
\author{Yonglun Jiang}
\author{Eric R. Weeks}
\affiliation{Department of Physics, Emory University, 400 Dowman Dr., Atlanta, GA 30322, USA.}
\author{Nicholas P. Bailey}
\affiliation{ ``Glass and Time'', IMFUFA, Dept. of Science and Environment, Roskilde
 University, P. O. Box 260, DK-4000 Roskilde, Denmark}

\begin{abstract}
  We have studied shear deformation of binary Lennard-Jones glasses to investigate the extent to which the transient part of the stress strain curves is invariant when the thermodynamic state point is varied along an isomorph. Shear deformations were carried out on glass samples of varying stability, determined by cooling rate, and at varying strain rates, at state points deep in the glass. Density changes up to and exceeding a factor of two were made. We investigated several different methods for generating isomorphs but none of the previously developed methods could generate sufficiently precise isomorphs given the large density changes and non-equilibrium situation. Instead, the temperatures for these higher densities were chosen to give state points isomorphic to the starting state point by requiring the steady state flow stress for isomorphic state points to be invariant in reduced units. In contrast to the steady state flow stress, we find that the peak stress on the stress strain curve is not invariant.  The peak stress decreases by a few percent for each ten percent increase in density, although the differences decrease with increasing density. Analysis of strain profiles and non-affine motion during the transient phase suggests that the root of the changes in peak stress is a varying tendency to form shear bands, with the largest tendency occurring at the lowest densities. We suggest that this reflects the effective steepness of the potential; a higher effective steepness gives a greater tendency to form shear bands.
\end{abstract}

\maketitle

\section{Introduction}


In recent years it has been realized that many model systems for simulating liquids and glasses have a hidden scale invariance, whereby curves in the phase diagram can be identified along which many structural and dynamical properties are invariant when expressed in an appropriate scaled-unit system. These curves are called isomorphs \cite{Gnan2009}. Isomorphs have been studied extensively using computer simulations \cite{Bailey/others:2008b,Bailey/others:2008c,Schroder/others:2009b,Gnan2009,Schroder/others:2011} of many different model systems \cite{Veldhorst/Dyre/Schroder:2014,Hummel/others:2015}, and experimental consequences have also been tested \cite{Gundermann/others:2011, Hansen2018}. Reviews of the overall theoretical framework and the many interesting consequences arising from its basic assumptions can be found in Refs.~\onlinecite{Pedersen/others:2011,Dyre:2014,Dyre:2016}. An example of the use of isomorph concept as a theoretical tool is a method for efficient calculation of melting curves \cite{Pedersen/others:2016a,Singh/Dyre/Pedersen:2021}. While exact isomorphs exist only in certain model systems, they can nevertheless help to explain a great deal of the behavior of realistic systems.

More recently, the consequences of hidden scale invariance in non-equilibrium situations, especially aging, have begun to be studied \cite{Dyre2020}. An important class of non-equilibrium phenomena involves shear deformation and plastic flow. The first study of isomorphs in a sheared system was already in 2013 \cite{Separdar2013}, where the single-component Lennard-Jones fluid and the Kob-Andersen binary fluid\cite{Kob/Andersen:1994,Kob/Andersen:1995a,Kob/Andersen:1995b} were studied in planar Couette flow in steady-state conditions. In 2019 we published work studying deformation of Kob-Andersen glasses under steady state flow with relatively modest density changes, up to 10\% \cite{Jiang2019}. There we showed that the statistics of the steady state rheology are isomorph invariant: the flow stress, its fluctuations and autocorrelation, as well as distributions of stress changes over small strain intervals, at varying strain rates.

The aim of the present work is to study isomorphs in a true out-of-equilibrium context, focussing on the transient behavior of sheared glasses, specifically the initial part of the stress strain curve, characterized by an (approximately) linear stress rise corresponding to (approximately) elastic behavior, followed by a stress peak, and then relaxation towards the steady state. The transition from non-flowing to a flowing state represents complex, non-linear behavior, while the steady state flow is also non-linear in that it exhibits shear-thinning (the flow stress rises more slowly than linearly with strain rate), as discussed in Appendix \ref{rheology}. Plasticity of glasses is thus both highly non-linear and highly non-equilibrium and thus offers a stringent test of isomorph invariance. Isomorph invariance of linear transport coefficients has already been investigated in detail, at least in the liquid state, see for example Refs.~\onlinecite{Heyes/Others:2019} and \onlinecite{Knudsen/Others:2021}. We are interested in to what extent the stress-strain curve collapses along a given isomorph (with given cooling and strain rates), and will focus particularly on the peak stress in the transient phase, and the flow stress for comparison. These two quantities (in reduced form) are convenient to plot as a function of density along isomorphs, giving a quick overview of the degree of invariance. We are also interested in attempting larger density changes than before. The model studied is the usual Kob-Andersen binary Lennard-Jones system \cite{Kob/Andersen:1994,Kob/Andersen:1995a,Kob/Andersen:1995b}.

For the prior work on the steady-state behavior the main thermodynamic parameters were density, temperature and strain rate. Since the first two were linked along an isomorph, there were effectively two parameters: a parameter labelling the isomorph (in principle the excess entropy\cite{Gnan2009}), and the reduced-unit strain rate. A feature of the transient state regime is that the thermal history of the glass prior to deformation becomes relevant. Our glasses are prepared by cooling at a fixed cooling rate, from a temperature near the melting point at the lowest density considered. Each deformation simulation is thus characterized by four parameters: the density $\rho$, temperature $T$, cooling rate $R_{c,0}$ and strain rate $\dot\epsilon$. Since we focus on trying to identify isomorphs, density and temperature are varied according to the (putative) isomorph, and cooling rate always refers to the initial (low) density in real units (hence the subscript 0), while strain rate is referred to using {\em reduced units} (see Sec~\ref{sec:reduced_units}; isomorph invariance can only be expected when the strain rate is fixed in reduced units \cite{Separdar2013}). For shearing simulations at densities higher than that of the cooling runs, configurations were scaled uniformly to the desired density (after cooling but before shearing) and the kinetic energy set to that appropriate for the corresponding (isomorphic) temperature. In this way the thermal histories of starting configurations at different points along the isomorph are identical by construction.

Our primary interest is to determine whether the stress-stress curve in its entirety is 
invariant along isomorphs. A crucial part of this is the question of how to determine isomorphs in non-equilibrium situations:  that is, given an initial temperature and density $(T_i,\rho_i)$, we need to find a new temperature $T_f$ for a desired $\rho_f$ for which the system properties are isomorphic.  There are several ways identified in prior work to find $T_f$, and here we show that these methods all identify similar but nonetheless different values for $T_f$ when the density change ($\rho_f/\rho_i$) is large.  More troubling, these methods lead to slightly different values of the (reduced) flow stress and peak stress, that is, these quantities are not isomorphically invariant using these prior methods for finding $T_f$.  In order to proceed we therefore adopt a pragmatic approach and identify an isomorph candidate by requiring the flow stress to be invariant.  The temperature $T_f$ thus identified falls in the middle of those predicted by the prior methods.  Having an isomorph candidate defined this way we then investigate the full stress-strain curves, and find that an invariant flow stress does not ensure that the whole curve is invariant: the peak stress tends to decrease with increasing density along the isomorph candidate. This is our main result: that there do not exist $\rho, T$ curves along which the reduced stress-strain curves are invariant.  Analysis of the particle motions suggests a cause: a tendency to develop shear bands despite the use of a SLLOD algorithm which favors homogeneous flow.  This tendency varies according to density.

\section{\label{sec:reduced_units}The isomorph approach}

The heart of the existence of isomorphs is that a phase space trajectory at one density and temperature can be scaled to another density (corresponding to a scaling of space) and another temperature (corresponding to a scaling of time, and thereby velocities and kinetic energy) and be, in fact, a valid trajectory at the new state point. Alternatively, trajectories can be found at the two isomorphic state points which are identical apart from rescaling space and time, and equally probable in their respective ensembles. This means considering the reduced position coordinates $\tilde{\mathbf{r}}_i\equiv \rho^{1/3} \bfa{r_i}$; {\it i.e.}, scaling essentially by the average interparticle spacing, using $\rho = N/V$, the number density of the system with $N$ particles in a volume $V$.  The reduced time is $\tilde t = t \rho^{1/3}\sqrt{k_B T/\langle m\rangle}$; {\it i.e.}, scaling essentially by the time for a particle with the mean mass $\langle m\rangle$ to cross an interparticle spacing with the thermal velocity. This is sometimes called the ``same movie" principle \cite{Dyre:2018b}. It follows that a correct comparison of isomorphic trajectories involves putting all quantities into dimensionless form, called ``putting into reduced units," by scaling lengths and times as above, and consistent with these, energies by $k_BT$. Masses are simply scaled by the average particle mass (a non-dynamical, non-thermodynamic quantity, which in our model is set to unity). This unit system was introduced by Rosenfeld \cite{Rosenfeld1977}; the scaling for all other quantities can be derived from these \cite{Gnan2009}. As an example the flow stress, having units of energy density, has the reduced form

\begin{equation}\label{eq:sigma_tilde}
    \tilde \sigma_f \equiv \frac{\sigma_f}{\rho k_B T}.
\end{equation}
When we talk of ``invariant flow stress" or small deviations therefrom, it must be remembered that we refer always to the dimensionless, reduced-unit form. The real flow stress in our simulations varies by over a factor of a hundred (2 from the density and 50 from the temperature).

In practice, it is well known that isomorphic invariance is imperfect in many situations \cite{Dyre:2014}.  Inverse power law (IPL) systems are notable for which isomorphic curves $(\rho,T)$ can be analytically calculated and the isomorphic properties are exact; here the phase space trajectories are mathematically identical along the isomorph curves.  However, in systems like the Kob-Andersen binary Lennard-Jones system that we study, isomorphic invariance is not exact.  As one moves along the isomorph curve $\rho,T$, even in reduced units the peak height of the pair correlation function $g(r)$ could vary by 1-2\%, for example.  The definition of a successful isomorph, then, is similar to many physics theories and approximations:  while not exact, one wants the isomorph description to ``explain'' most of the observed behavior. Rather than needing to understand a system at all values of $\rho,T$, one could know the properties of a system at a given $\rho$ and a variety of $T$ and then know, to some degree of confidence, what happens at other $\rho$ and isomorphically matched $T$.  As a richer example, consider the aging of glassy materials.  Aging occurs when an equilibrated liquid is quenched (by decreasing $T$, increasing $\rho$, or both) to a nonequilibrium glassy state.  As the state is out of equilibrium, the sample properties slowly evolve with time.  If one can identify isomorphically matched states in the glassy regime, then a simulation of a sample temperature-quenched at constant density could predict the aging behavior of a sample density-quenched at constant temperature, as long as the final state is on the same isomorph $\rho,T$ curve \cite{Gnan/Others:2010}.

Thus, a successful isomorph strategy will identify a curve $\rho,T$ over which all sample properties of interest (in reduced units) are similar, perhaps with variations of no more than a few percent.  In this work, our properties of interest are the stress-strain curves of glassy samples deeply out of equilibrium.  We demonstrate that it is impossible to match both the peak stress and steady-state flow stress in this isomorphic sense.  In particular, matching the steady-state flow stress (variations less than one percent) results in the peak stress varying by about 20\% from the lowest to highest densities studied.

To end this section, we point out that isomorph invariance is distinct from the concept of time-temperature superposition, according to which certain glass-forming liquids have frequency-dependent responses whose spectral shape is invariant\cite{Harrison:1976,Richert:2005}. Isomorph invariance is both a weaker claim, in that it relates quantities along a given isomorph and not throughout the phase diagram, and a stronger claim, in that it specifies how the overall timescale varies along an isomorph, which TTS does not. The relation between the two has been discussed in Ref.~\onlinecite{Niss/Hecksher:2018} (who refer to ``isochronal scaling", rather than ``isomorph invariance").

\section{Model, starting parameters, basic procedures}

\begin{figure}[ht]
  \includegraphics[width=0.48\textwidth]{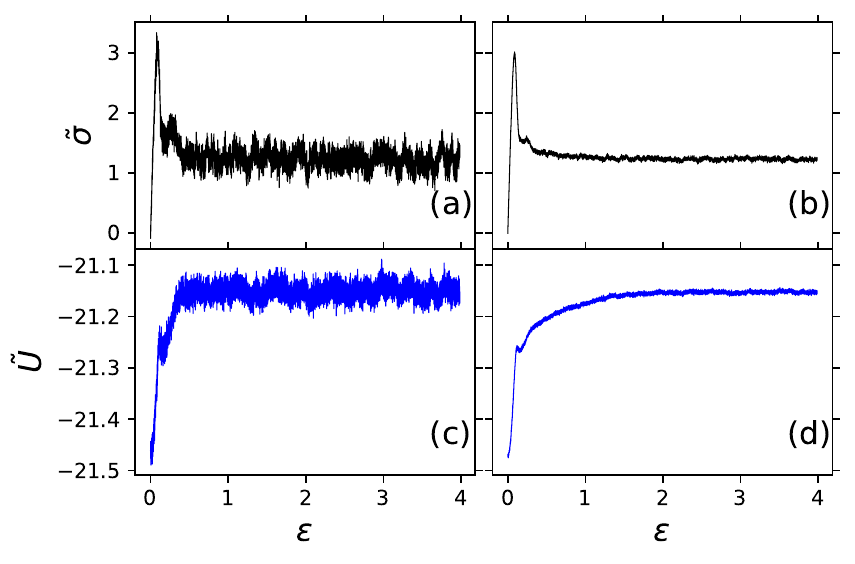}
  \caption{Examples of reduced-unit shear stress (top panels) and potential energy per particle (bottom panels) versus strain, from shear deformation of a binary Lennard-Jones glass with density 1.183 at temperature 0.3. The glass was prepared by cooling at the rate $10^{-7}$ (LJ units), while the deformation was carried out at reduced strain rate $10^{-5}$, meaning the real strain rate was $10^{-5}\rho^{1/3}(T)^{1/2}\simeq 5.8\times 10^{-6}$. The left panels show curves from a single run, while the right panels show the average of 40 independent runs.  }
  \label{fig:sspe_curves}
\end{figure}

In this section we give a brief overview of the model we use, how we cool, and how we deform. We work with the standard Kob-Andersen 80-20 binary Lennard-Jones system with the only slight difference being that we use a shifted-force cutoff \cite{Toxvaerd2011}. This is imposed at distance 2.5 $\sigma$ for each interaction type. When expressing in ordinary (non-reduced) units we use the unit system defined by the energy parameter of the AA interaction, $\epsilon_{AA}$, the length parameter of the AA interaction, $\sigma_{AA}$ and the particle mass (common to both A and B particles). For brevity we refer to these as ``Lennard-Jones" units. A challenge with investigating the transient part of the stress and strain curve is that the fluctuations cannot just be averaged out by running longer. Consider the example shown in Fig.~\ref{fig:sspe_curves}; the fluctuations in panel (a) are quite severe and avalanche-like, even though the temperature is not as low as in our last paper \cite{Jiang2019}. To manage the fluctuations we use fairly large system sizes ($N=10000$) and multiple independent runs, starting from an ensemble of configurations generated by separate cooling runs. There are 40 members in this ensemble. Studying the run-to-run deviations also allows us to determine errors on flow stress and peak stress quite precisely.

\subsection{Cooling}

The lowest density we consider is 1.183, close to the density 1.2 studied by Kob and Andersen and most often by others. At this density we run a liquid simulation involving $N=10000$ particles at the relatively high temperature of 1.0 (close to the melting temperature for this composition\cite{Pedersen2018}). Forty independent configurations were sampled from this NVT run and used as the start of the cooling runs. We have cooled at rates $10^{-5}$, $10^{-6}$, and $10^{-7}$ in Lennard-Jones units down to temperature 0.3, deep in the glassy state.  The cooling is done at constant density.

\subsection{Shear deformation}
\label{subsec:sheardeform}

For shear deformation simulations we employ the SLLOD\cite{Evans/Morriss:1984,Ladd:1984} equations of motion and Lees-Edwards boundary conditions \cite{Lees/Edwards:1972,Allen/Tildesley:1987}. The direction of shearing is the $x$-direction with the gradient in the $y$-direction, therefore the relevant component of stress is $xy$. As is common in SLLOD-based simulations the temperature is controlled using an isokinetic thermostat, which maintains a fixed kinetic energy, chosen\footnote{Because of the constraint of conserved kinetic energy, the number of degrees of freedom is actually $3N-4$, not $3N-3$ \cite{Morriss/Dettmann:1998}. Thus we actually set the kinetic energy slightly too high, by 1 part in 30000; our results apply to temperatures which are higher than the stated temperatures by this amount.} to be $3/2(N-1) k_BT$.  We choose the shear rate to be fixed in reduced units \cite{Separdar2013} to the value $10^{-3}$, $10^{-4}$,  or $10^{-5}$. We also choose the time step to be fixed in reduced units, which is practical--it automatically ensures that a time step which is stable at one density and temperature will be stable along the isomorph. We choose a real time step of 0.004 (LJ units) at the lowest density 1.183 and temperature 0.3; thus the real time step is smaller at higher densities and temperatures, proportional to $\rho^{-1/3}T^{-1/2}$. Note, the Reynolds number of the flow in our simulations is extremely low, of order $5\times10^{-4}$ for our highest strain rate. Thus problems associated with highly sheared liquids, such as non-linear profiles and string phases\cite{Evans/Morriss:1986}, non-unique or anisotropic temperature\cite{Baranyai:2000}, and anisotropic stress\cite{Hoover/Hoover/Petravic:2008}, are not relevant here and more advanced techniques such as profile-unbiased thermostats \cite{Evans/Morriss:1986,Morriss/Dettmann:1998} are not required. \newpurple{As an extra check, though, we have repeated some of our simulations with a modified thermostat which does not couple to the velocities in the flow ($x$) direction, and found essentially identical results.}

In the following it turns out best to use the flow stress, defined as the mean stress during the steady state regime, to determine isomorphs. For this purpose we define strains greater than 2 (200\%) as the steady state for all shear simulations. This choice originates from inspection of the potential energy versus strain curve; we find that the potential energy of the slowest cooled configuration under slowest shear gradually reaches the steady state value at a strain between 1 and 2, see Fig.~\ref{fig:sspe_curves} (and also Fig.~\ref{fig:pe_curves}), which shows examples of stress-strain curves and potential energy-strain curves. In this case the cooling rate for the glasses was 10$^{-7}$ in LJ units and the reduced shear rate was $10^{-5}$, corresponding to a real shear rate of about $5.8 \times 10^{-6}$. By eye the shear stress seems to be essentially at its steady-state value by strain 0.5, but for this slowly cooled system the potential energy has not converged to the steady-state value until around strain 2.0\cite{Singh2020}. 

We need to analyze the stress and strain data to extract mean values as well as uncertainties, so it is worth briefly giving the details of these calculations.  Given the many independent runs, the total strain for each shearing run can be relatively modest, specifically 4 (400\%), of which the last 200\% is used for determining the flow stress.  We find the mean reduced stress in this steady state strain regime for each shear simulation, giving 40 independent estimates. We then use the average of 40 shear runs as the flow stress at the corresponding density, temperature, strain rate, and cooling rate.  Error bars are computed using the usual formula for the standard error on the mean\cite{Taylor:1996}:  dividing the sample standard deviation of the 40 data points by $\sqrt{40}$. For the stress peak height (preceding the steady state), each individual stress-strain curve is too noisy to determine accurate values; accordingly, we take the 40 independent runs and average them 8 at a time to give five independent stress-strain curves.  We then find the peak stress height for one curve by fitting the region of the averaged stress-strain curve around the peak to a fourth-degree polynomial. The interval for fitting is the strain with the numerically largest shear stress, plus or minus 0.05.  Taking these five groups each averaged over 8 simulations, we average those five stress peak values to define the measured peak stress height.  The uncertainty of this value is then the sample standard deviation of the 5 estimates divided by $\sqrt{5}$.

\section{\label{sec:identify_iso_noneq}Identifying candidate isomorphs}

Determining isomorphs, or candidates for isomorphs, in non-equilibrium situations, is a crucial task and the subject of the section. Dyre has presented a general framework extending isomorph theory to non-equilibrium situations \cite{Dyre2020}. The first ingredient is the concept of systemic temperature, $T_s$ which can be defined for an individual configuration. Given the potential energy of that configuration, $T_s$ is the temperature at which (for the same density) the equilibrium potential energy is that configuration's potential energy. Given a change in density, and assuming perfect hidden scale invariance, the dynamics will be invariant as long as the ratio of the bath (i.e. thermostat) temperature to the system temperature is the same. Thus the question of identifying the correct bath temperature at the new density is the question of determining by what factor the systemic temperature of the initial configuration changes when its density is scaled uniformly.

The definition is not a practical way to determine $T_s$ in a glassy system because determining the equilibrium potential energy as a function of temperature is not feasible. Instead we turn to another recent work, by Schr{\o}der, who has developed a method for predicting isomorphic tempratures by comparing the forces on particles in a configuration before and after uniform scaling. This method gives a temperature ratio corresponding to a change of density of a single configuration from $\rho_i$ to $\rho_f$ as

\begin{equation}\label{eq:force_method}
    \frac{T_f}{T_i} = \left(\frac{\rho_i}{\rho_f}\right)^{1/3} \frac{|\bfa{F}_f|}{|\bfa{F}_i|}
\end{equation}
Moreover we can argue (see appendix \ref{sec:force-method-systemic-temperature}) that this corresponds to the change in systemic temperature, and therefore is the factor by which the bath temperature should be changed.

In the rest of this section we determine candidate isomorphic temperatures using the force method and other methods based on those used for equilibrium isomorphs. We show that none of them yield invariant flow stresses, and instead we generate a candidate isomorph by constructing it to have an invariant flow stress. The starting point in all cases is a shearing simulation with a cooling rate $R_{c,0}=10^{-5}$  and (reduced) strain rate $10^{-3}$, at the lowest density (1.183), which is referred to as the reference density.

\subsection{Force method and fluctuation methods}

In this subsection we show that the force method for generating candidate isomorphs, along with other fluctuation methods, give different temperatures from those required for invariant flow stress. Different methods disagree by increasingly large amounts as the density factor increases, although the rate of increase decreases with increasing initial density. To compare different methods, we use as a test case glasses cooled at $R_{c,0}=10^{-5}$ (LJ units) and sheared at $\tilde{\dot{\epsilon}}=10^{-3}$ (reduced units), and consider mostly the case of a density increase of 10\% from 1.183 to 1.301. The results for the temperature ratio ($T_f/T_i$) and corresponding estimated temperature at the latter density are summarized in Table~\ref{tab:compare_methods_density1}. Since the force method only requires a single configuration to give a temperature ratio $T_f/T_i$, it can be applied repeatedly throughout a simulation at the reference density, giving an immediate estimate of statistical errors and possible systematic changes (for example as a function of strain).

\begin{figure}[htp]
    \centering
    \includegraphics[width=0.48\textwidth]{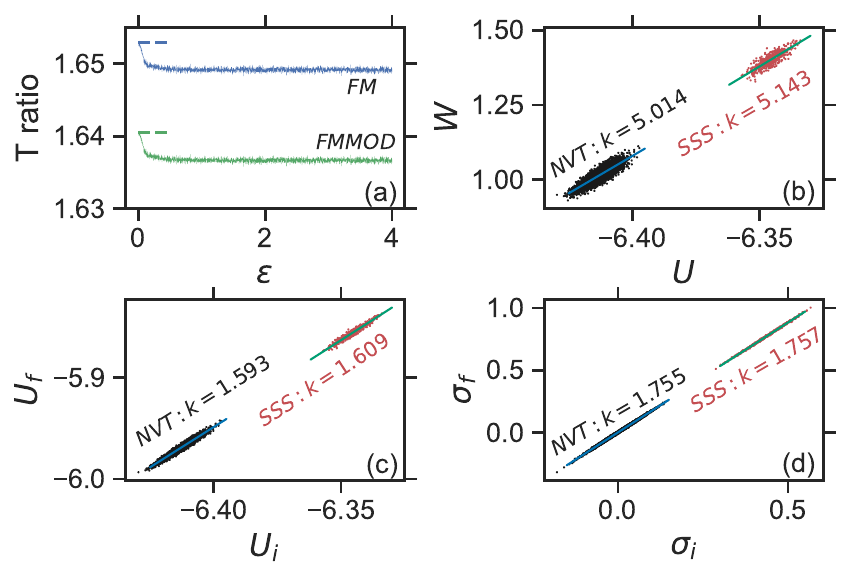}
    \caption{The five methods to obtain the temperature ratio $T_f/T_i$ used to identify the state point $\rho_f, T_f$ isomorphic to state point $\rho_i, T_i$ from simulations at the latter. The example here uses reference density and temperature $\rho_i = 1.183$ and $T_i=0.3$ and a starting configuration cooled at $R_{c,0}=10^{-5}$. We consider a new density $\rho_f = 1.1\rho_i=1.301$. Black indicates results from NVT simulations and red is for the steady state from a shear simulation with the highest (reduced) strain rate $\tilde{\dot{\epsilon}}=10^{-3}$. For (a), (c) and (d) the system was sampled at regular intervals during the simulation and configurations uniformly scaled to $\rho_f$; the potential energy, forces, and shear stress were calculated on the scaled configurations, denoted with subscript $f$. (a) Temperature ratio given by the force method FM (blue), Eq.~\eqref{eq:force_method}, and modified force method (green), Eq.~\eqref{eq:modified_force_method} from the same shear simulation. The dotted horizontal lines indicate the corresponding temperature ratios from the NVT simulation. (b) Scatter-plot of the virial $W$ versus potential energy $U$. The slopes (correlation coefficient) of the two fits are $5.014$ ($0.869$) and $5.143$ ($0.859$) respectively, where the slopes can be considered estimates of the density scaling exponent $\gamma$, which yields the temperature factor via Eqs.~\eqref{eq:isomorph_from_h_rho} and \eqref{eq:B_A_from_gamma}. (c) Scatter-plot of $U_f$ against $U_i$  (DIC-pe). The slopes (correlation coefficient) are $1.593$ ($0.975$) and $1.609$ ($0.972$) respectively. Here the slopes correspond directly to the temperature ratios. (d) Scatter-plot of $\sigma_{f}$ against $\sigma_{i}$. The slopes (correlation coefficient) are $1.755$ ($1.00$) and $1.757$ ($0.999$). Here the temperature ratio is the slope divided by the density ratio (1.1). Table~\ref{tab:compare_methods_density1} gives the results of the different methods.}
    \label{fig:4methods}
\end{figure}

Figure~\ref{fig:4methods}(a) shows the temperature ratio for the force method, along with that from a modified version

\begin{equation}\label{eq:modified_force_method}
    T_f =  T_i \left(\frac{\rho_i}{\rho_f} \right)^{1/3} \frac{\bfa{F_i}\cdot \bfa{F_f}}{\bfa{F_i}\cdot \bfa{F_i}}
\end{equation}
The modified force method is based on simple linear regression of the scaled forces as a function of the unscaled forces, and gives a slightly lower temperature estimate. In the figure, there is a clear but small systematic decrease in the estimated temperature ratio, (about 0.25\% and 0.6\% for the force method and its modification, respectively, as the strain increases from zero to steady state conditions). It is also clear that the statistical fluctuations are rather small in the steady state, so that the estimate from a single configuration would indeed give a precise estimate of the ensemble average. The estimate from the modified force method is systematically lower by about 0.7\% in the steady-state.

\begin{table}
    \centering
    \begin{tabular}{c|c|c|c}
    Method & Ensemble & $T$-ratio & $T_f$ \\
    \hline
    FM & NVT & 1.653 & 0.4959 \\
    FM & SSS & 1.649 & 0.4947 \\
    FM-mod & NVT & 1.647 & 0.4940 \\
    FM-mod & SSS & 1.637 & 0.4911\\
    $WU$/analytic & NVT & 1.593 & 0.4779 \\
    $WU$/analytic & SSS & 1.609 & 0.4828 \\
    DIC-pe & NVT & 1.593 & 0.4779\\
    DIC-pe & SSS & 1.609 & 0.4827 \\
    DIC-sts & NVT & 1.595 & 0.4786 \\
    DIC-sts & SSS & 1.597 & 0.4791 \\
 
    \hline
    Matching flow stress & SSS & 1.623 & 0.4869 \\
    \end{tabular}
    \caption{Comparison of methods for identifying isomorphic temperature upon raising density by 10\% from $\rho_i=$1.183 to $\rho_f=$1.301, for glasses cooled at rate $10^{-5}$ to $T_i = 0.3$. SSS refers to steady state shearing at $\rho_i, T_i$, data taken between strains 2 and 4, with reduced strain rate $10^{-3}$; NVT refers to NVT simulations of $10^7$ steps at $\rho_i, T_i$. The temperature ratio that best matches the flow stress is $1.623$, listed in the last line of the table.}
    \label{tab:compare_methods_density1}
\end{table}

The other panels in Figure~\ref{fig:4methods} illustrate estimation of isomorphic temperatures by other methods, based on fluctuations in both NVT (unstrained) simulations, and the steady state part of sheared simulations (SSS). Fig.~\ref{fig:4methods}(b) shows  $U,W$ fluctuations, both NVT and SSS, from which a density scaling exponent $\gamma$ can be found as the regression slope. For Lennard-Jones potentials an analytic expression for the shape of isomorphs can be found; as detailed in Appendix~\ref{subsec:analytic}, the single free parameter can be fixed using the measured $\gamma$, allowing the temperature ratio for a given density change to be determined.  We show data for a glass cooled at the fastest rate; using the slowest cooled glasses gives a difference less than 0.1\% in the resulting temperature ratio. We refer to this method in table and figure legends as $UW$/analytic. The NVT value for $\gamma$ is 2.5\% lower than the value from the steady state fluctuations, giving a temperature ratio 1\% lower for the 10\% density increase; the difference will increase with larger density jumps.

Fig.~\ref{fig:4methods}(c) shows the direct isomorph check (DIC) whereby potential energies from scaled configurations are plotted against those from unscaled configurations (i.e. drawn from the simulation at the reference density). Determining the slope gives a direct estimate of the temperature factor, 1.609 from the steady-state and 1.593 in NVT.  These are equal to the estimates from  $UW$/analytic, consistent with $\gamma$ being essentially the DIC in the limit of infinitesimal density changes\cite{Gnan2009}. Fig.~\ref{fig:4methods}(d) shows a DIC-like method based on shear-stress fluctuations. The observed slope when plotting the scaled versus unscaled shear stresses is not the temperature ratio, but includes also a factor of the density ratio $\rho_f/\rho_i=1.1$. After dividing the latter out, the temperature ratio estimate from steady state fluctuations is slightly lower (0.7\%) than the corresponding energy-based DIC estimate, see Table~\ref{tab:compare_methods_density1}. Interestingly, the correlation is much higher for the stress-based DIC than for the energy-based DIC, and the difference between NVT and SSS estimates is much smaller than for the energy-based DIC, at only 0.1\%. These two estimates are also very close to the NVT energy-based DIC estimate.

\begin{figure}[htp]
    \centering
    \includegraphics[width=0.48\textwidth]{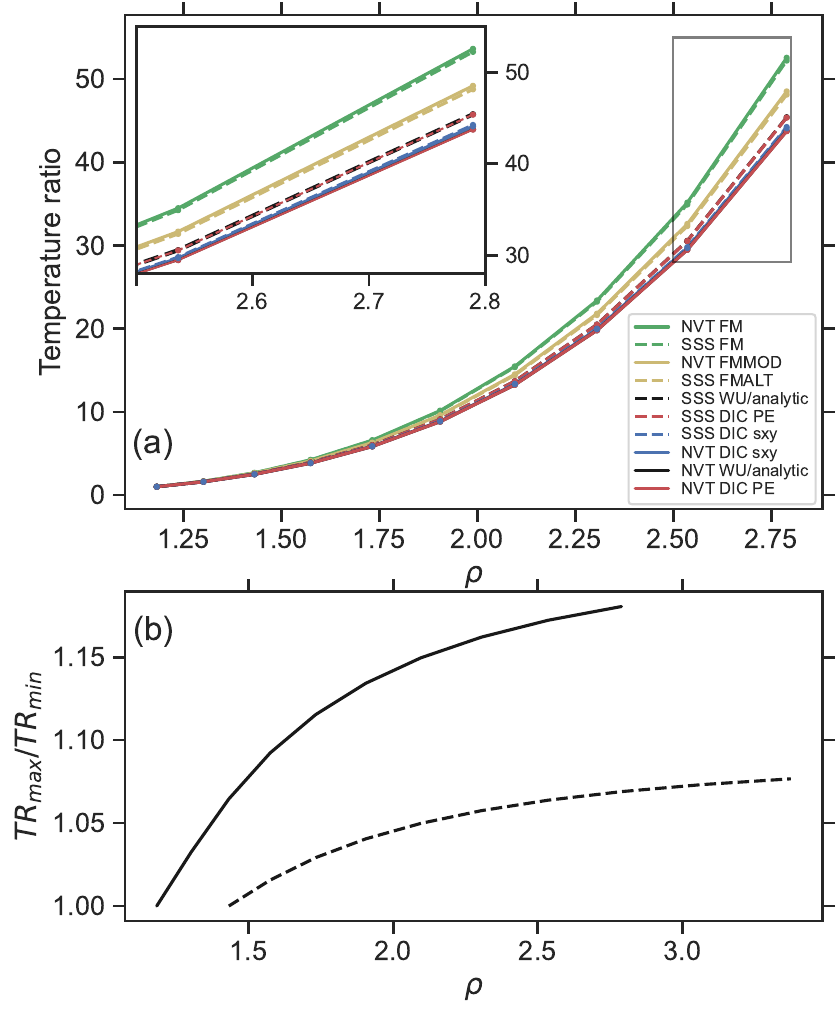}
    \caption{Top panel: temperature ratios from different methods against density for starting density $1.183$. Solid lines are for NVT simulation (NVT) $10^7$ steps and dashed lines are for steady state (SSS). Inset zooms in for the largest two densities.  Bottom panel: the ratio between the highest and lowest temperature factors of the SSS from the top panel versus density. Solid line starts at $\rho=1.183$, and dashed line starts at $\rho=1.301$.}
    \label{fig:different_t}
\end{figure}

To summarize the above results, the largest temperature factor is given by the force method in the sheared (NVT) system, while the smallest is given by either the $\gamma$ method or DIC using NVT data or the stress-based DIC using either NVT or SSS data. The spread between highest and lowest SSS values is 3.5\%. We next consider how this variation depends on the size of the density jump. Fig. \ref{fig:different_t}(a) shows the temperature ratios from the aforementioned methods versus density. For small density changes, all these methods return similar results with little discrepancy. This discrepancy increases with larger density spans, indicating the challenge of identifying isomorphs in these situations. Other cooling rates and strain rates were also checked and found to return almost the same results, in particular the order of temperature-estimates is identical. On the other hand, the same calculation of temperature ratios (TR) but at a 10\% higher starting density shows much less difference between various methods, as shown in Fig. \ref{fig:different_t}(b). The virial-potential energy correlation coefficient $R$ gives a general indication of the quality of hidden scale invariance, with 0.9 being a conventional criterion for good isomorphs\cite{Gnan2009}; for our lowest density $R$ is lower than is, around 0.86-0.87 (Fig.~\ref{fig:4methods}(b)), therefore it is not surprising that the methods give diverging estimates. For the next lowest density, 1.301, the value of $R$ is 0.972, substantially higher, 
and therefore one can expect less divergence between the different methods, as the potential energy landscape becomes increasingly well approximated by that of an inverse power law (IPL) particle interaction. 

\subsection{\label{sec:matching_flow_stress}Matching the flow stress}

\begin{figure}[htp]
    \centering
    \includegraphics[width=0.48\textwidth]{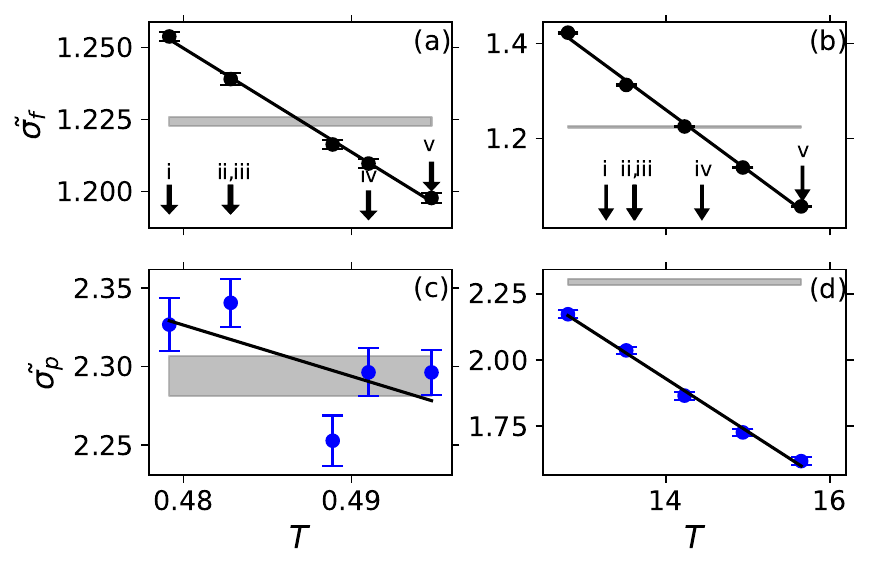}
    \caption{The reduced flow stress $\tilde{\sigma _{f}}$ and reduced peak stress $\tilde{\sigma _{p}}$ against temperature near the point of matching stress at $\rho_1 = 1.301$ (left two panels) and $\rho_9=2.789$ (right two panels). Glasses cooled (at lowest density) at rate $R_{c,0}=10^{-5}$  and sheared at reduced strain rate $\dot{\tilde{\epsilon}}=10^{-3}$. Solid black lines are linear fits. Gray horizontal lines indicate the $\tilde{\sigma _{f}}$ in panel (a) and (b), and $\tilde{\sigma _{p}}$ in (c) and (d) at the reference density with shaded region indicating error. Arrows in panel (a) and (b) point to $T$ estimated using: (i) DIC stress method; (ii) DIC PE method; (iii) $WU$/analytic; (iv) FMMOD; (v) FM.}
    \label{fig:fsxy_t}
\end{figure}

\begin{figure}[htp]
    \centering
    \includegraphics[width=0.48\textwidth]{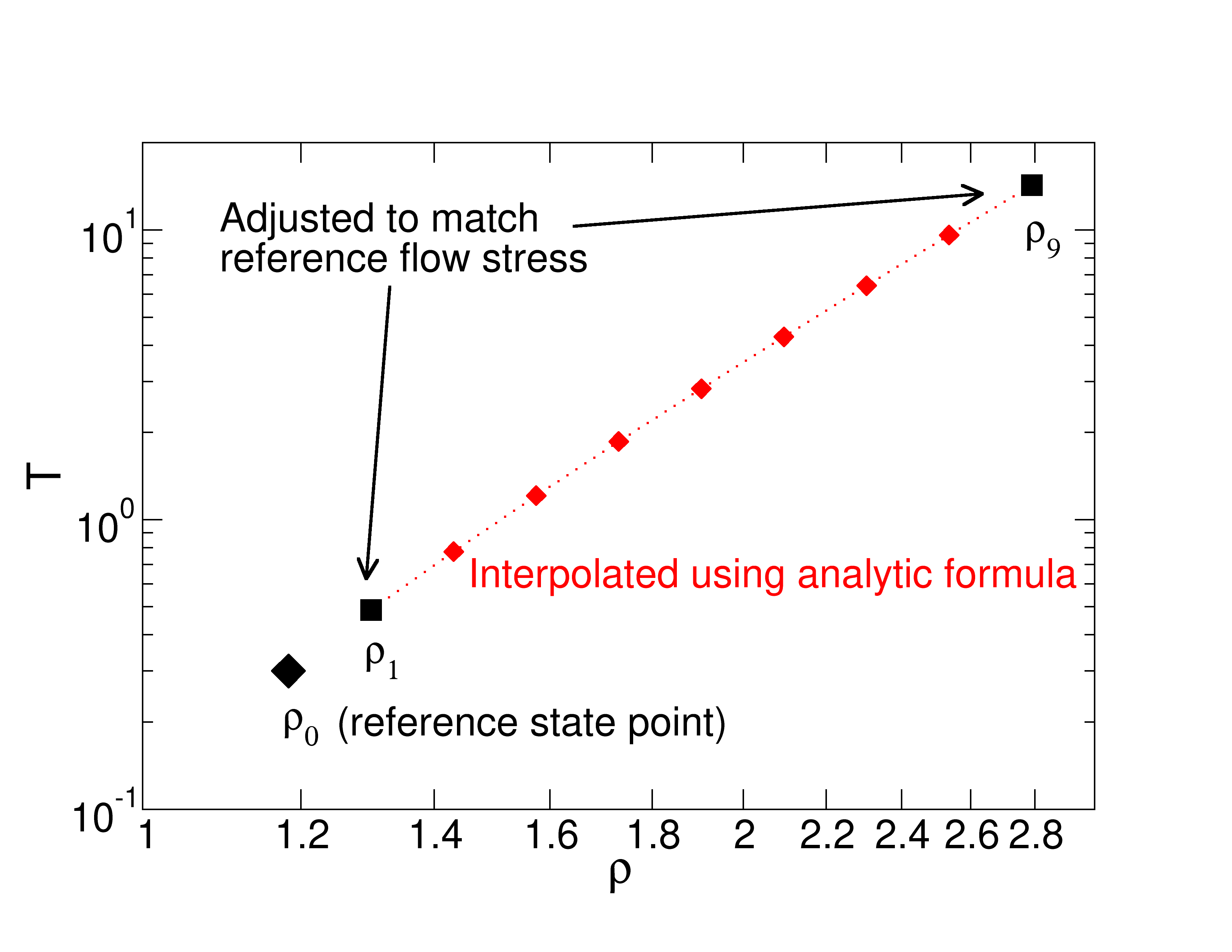}
    \caption{The isomorph determined by matching flow stress again on glasses cooled (at lowest density) at rate $R_{c,0}=10^{-5}$  and sheared at $\dot{\tilde{\epsilon}}=10^{-3}$. The large black diamond indicates the reference state point. The two black squares represent points whose temperatures were identified by matching the reduced flow stress to that of the reference, using the linear fits in Fig.~\ref{fig:fsxy_t}, and the red diamonds are points whose temperatures have been determined by interpolation between the black squares, using Eq.~\eqref{eq:isomorph_from_h_rho2}.}
    \label{fig:isomorph_illustration}
\end{figure}

Since the different methods yield a spread of potential isomorphic temperatures for a given density change, the question is then which of the above methods actually yields invariant quantities of interest. Considering the flow stress,   Fig.~\ref{fig:fsxy_t}(a) shows for density 1.301 the reduced flow stress  obtained by simulating at several temperatures, corresponding to some of the temperature estimates based on fluctuations. Panel (b) of the figure shows a similar plot for the highest simulated density; here the trial temperatures were not those given by the other methods, but chosen to span a similar range. The negative linear dependence of the reduced flow stress on temperature reflects that barriers to flow can be crossed more easily at higher temperature. The arrows indicate the temperatures determined by the various methods, which clearly do not match the reference flow stress within errors. The method that comes closest is the modified force method. The linear fits to the temperature dependence of the flow stress can, however, be used to accurately identify the temperature at which the $\tilde\sigma_f$ matches its reference value. 

\begin{figure}[htp]
    \includegraphics[width=0.48\textwidth]{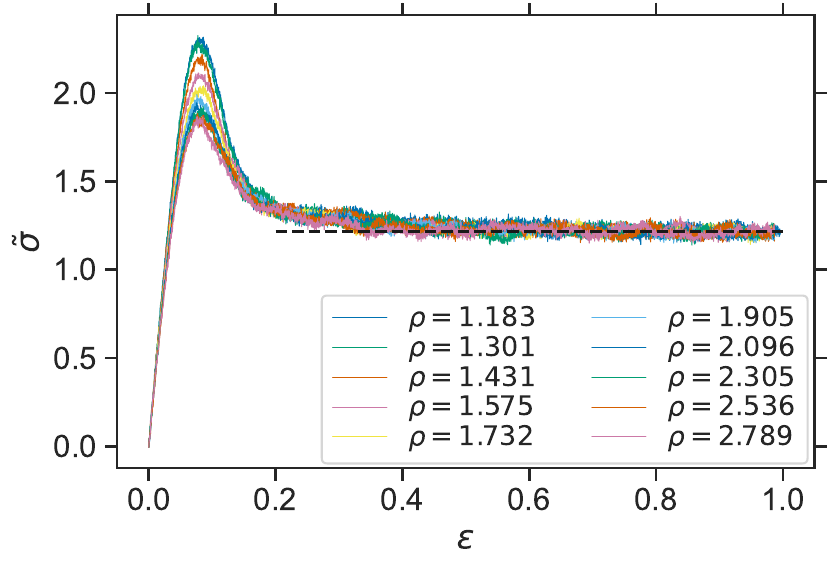}
    \caption{\label{fig:stress_strain_along_isomorph} Reduced stress-strain curves along isomorph determined by the procedure illustrated in Fig.~\ref{fig:isomorph_illustration}. Glasses cooled (at lowest density) at rate $R_{c,0}=10^{-5}$  and sheared at reduced strain rate $\dot{\tilde{\epsilon}}=10^{-3}$.}
  \end{figure}

Panels (c) and (d) of Fig.~\ref{fig:fsxy_t} show the measured reduced peak stresses at the same two densities and and the same temperatures as the flow stress in panels (a) and (b). Like the flow stress, the peak stress also has a negative linear correlation with temperature, although the larger errors combined with the limited temperature range reduce the apparent correlation in panel (c). Comparison with the value at the reference density indicated by the grey bars presages one of our main results, that for high density changes no temperature can be found which matches both the flow and peak stresses with their values at the reference state point. The main presentation of this result is in the next section (see Fig.~\ref{fig:stress_strain_along_isomorph}).

Our pragmatic approach to determining a candidate isomorph is to find the curve along which the reduced flow stress is invariant. This can always be determined, much like in equilibrium a curve of constant excess entropy can always be determined; further analysis then addresses the degree to which other quantities of interest are also invariant. The procedure of simulating several temperatures at each density in order to make a linear fit would be very time consuming if it should be done at each density of interest. To save work we can instead use the analytic expressions for Lennard-Jones isomorphs presented in Appendix~\ref{subsec:analytic}, Eqs.~\eqref{eq:isomorph_from_h_rho} and \eqref{eq:B_A_from_two_points}. For studying the isomorphs we consider from now on ten different densities, labelled with subscripts starting from zero, as $\rho_0=1.183$, $\rho_1=1.1\rho_0$, \ldots, $\rho_9=(1.1)^9\rho_0$. In applying the analytic expression for isomorphs, we have to treat $\rho_0$ separately. Related to its somewhat low value of the correlation coefficient $R$ (Fig.~\ref{fig:4methods}) and the fact that methods for determining isomorphic temperature starting from this density diverge rather quickly (Fig.~\ref{fig:different_t}), it turns out that no parameterization of the analytic formula can match the reduced flow stress over the full range from $\rho_0$ up to $\rho_9$. Such a parameterization can be found for the range $\rho_1$ to $\rho_9$, however. Thus we work as follows: Given the reduced flow stress at $\rho_0$ (and $T_0=0.3$) we use the linear fits in Fig.~\ref{fig:fsxy_t}(a,b) to determine the temperatures of matching reduced flow stress at densities $\rho_1$ and $\rho_9$, respectively. From Eq.~\eqref{eq:B_A_from_two_points} with $\rho_2$ replaced by $\rho_9$, and using these fit-determined temperatures $T_1$ and $T_9$ we fix the parameter $B/A$. Finally for the remaining densities $\rho_2, \ldots, \rho_8$ we determine the isomorphic temperatures  from Eq.~\eqref{eq:isomorph_from_h_rho} in the form

\begin{equation}\label{eq:isomorph_from_h_rho2}
    T_i = T_1 \frac{h(\rho_i)}{h(\rho_1)} = T_1 \frac{\rho_i^4 - (B/A) \rho_i^2}{\rho_1^4 - (B/A) \rho_1^2}.
\end{equation}

\begin{table}
    \centering
    \begin{tabular}{c|c}
    Density & Temperature \\
    \hline
    1.183 & 0.3\\
    1.301299 & 0.487031 \\
    1.431428 & 0.774540 \\
    1.574571 & 1.208396 \\
    1.732028 & 1.859225 \\
    1.905231 & 2.831010 \\
    2.095754 & 4.276669 \\
    2.305330 & 6.420943 \\
    2.535863 & 9.593855 \\
    2.789449 & 14.279829 \\
    \end{tabular}
    \caption{The densities and temperatures along the isomorph identified through matching the flow stress at the first, second, and last densities, and then using the analytical method to obtain the temperatures in between.}
    \label{tab:iso}
\end{table}

Figure~\ref{fig:isomorph_illustration} illustrates the construction of the isomorph. The resulting densities and temperatures are listed in Table~\ref{tab:iso}. These values were determined by matching the flow stress using the highest strain rate (reduced value $10^{-3}$) (and fastest cooling-rate, though that should not matter for the flow stress). Rather than separately repeat this procedure for identifying isomorphic temperatures for the other strain rates we take as a working hypothesis that the isomorphs in the $\rho-T$ plane determined by matching reduced flow stress do not depend on which reduced strain rate was used. This is consistent with empirical results of Separdar {\em et al.}\cite{Separdar2013} and theoretical arguments of Dyre\cite{Dyre2020}; its validity will be investigated in the following. The next step is the shearing deformation simulations. These are still very time-consuming, because at each density, and the corresponding temperature determined by the above procedure, 40 independent runs were carried out. This was repeated for all three cooling rates and all three strain rates. The stress-strain curves shown below are averages over the 40 runs in each case.

\section{\label{sec:stress_strain_curves}Analysis of stress-strain curves}

\begin{figure}[htp]
    \centering
    \includegraphics[width=0.48\textwidth]{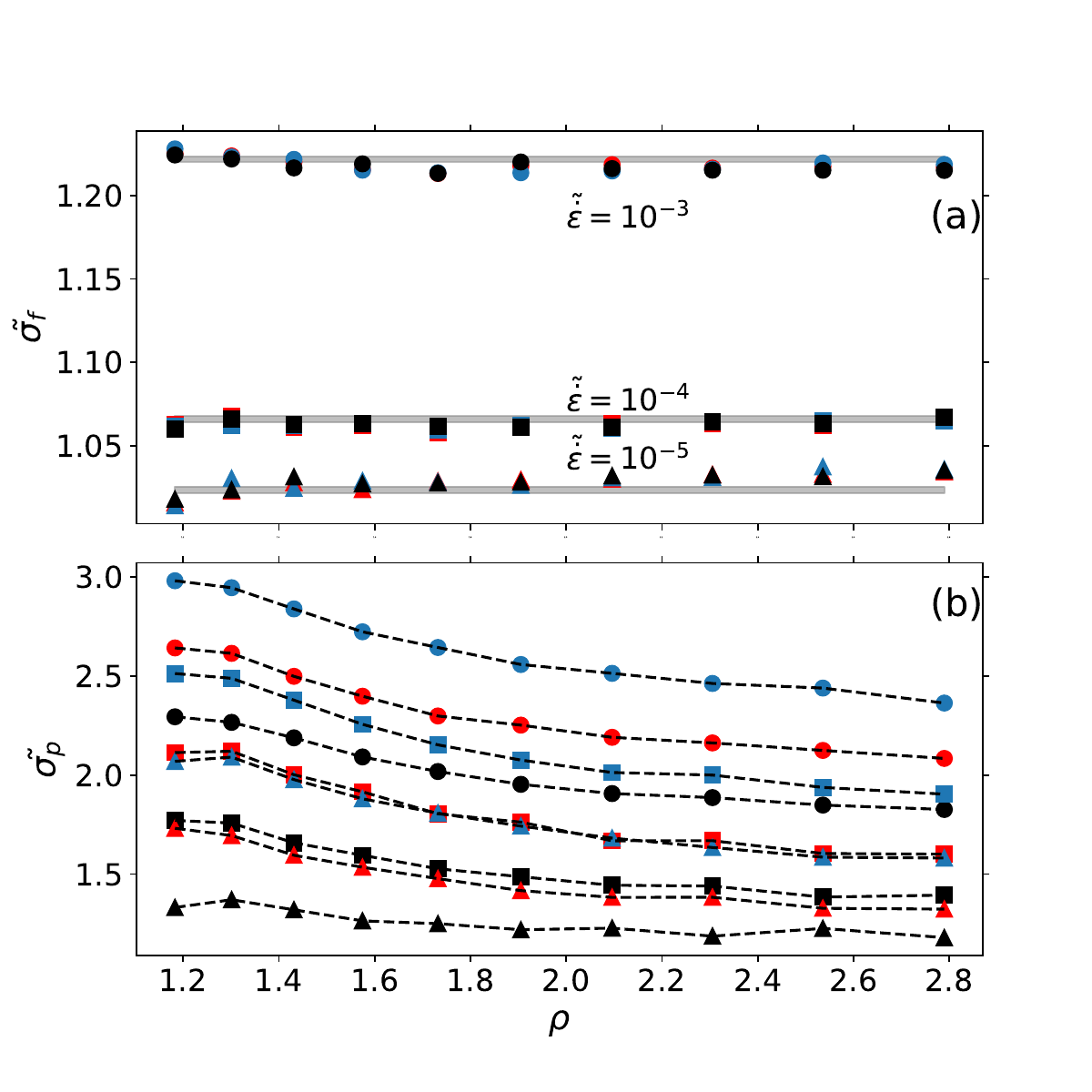}
\caption{(a) $\tilde{\sigma _{f}}$ and (b) $\tilde{\sigma _{p}}$ against density along the isomorph for $R_{c,0}=10^{-5}, 10^{-6}, 10^{-7}$ (black, red, blue), and $\tilde{\dot{\epsilon}}=10^{-3}, 10^{-4}, 10^{-5}$ (circle, square, triangle). Each point in (a) is an average of 40 shear simulations (on 40 individual configurations, or 30 for $\tilde{\dot{\epsilon}}=10^{-5}$) and the uncertainties are computed as described in Sec.~\ref{subsec:sheardeform}.  For the peak stress [panel (b)], we first divide the 40 shear runs into 5 groups and obtain 5 averaged stress and strain curves. We then fit the data for 10\% strain around the maximum stress using a fourth-degree polynomial and identify the maximum of the fit as the $\tilde{\sigma _{p}}$. The error is the standard deviation of the 5 values divided by $\sqrt{5}$. The three families in (a) correspond to $\tilde{\dot{\epsilon}}=10^{-3}, 10^{-4}, 10^{-5}$ from top to bottom respectively. The errors are all smaller than the marker size. The position of the gray bars in (a) are the reference $\tilde{\sigma _{f}}$ (at $\rho_1$ cooled with $R_{c,0}=10^{-5}$ and sheared with $\tilde{\dot{\epsilon}}=10^{-3}$) and the width of the bar indicates the reference $\tilde{\sigma _{f}}$ plus or minus the uncertainty.}
    \label{fig:flow_sts_along_isomorph}
\end{figure}

\subsection{Non-invariance of stress peaks}

With the protocol for determining the putative isomorph established, we now present our main results.
Figure~\ref{fig:stress_strain_along_isomorph} shows the full (reduced) stress strain curves for all densities along the isomorph generated as described in Sec.~\ref{sec:matching_flow_stress}, using reduced strain rate $10^{-3}$ and initial configurations cooled at rate $R_{c,0}=10^{-5}$. The initial elastic parts of the curves overlay, showing that the elastic shear modulus is invariant in reduced units along the isomorph. In particular, a closer inspection of the data (not shown here) shows collapse within the noise for the initial 1\% or so of strain. Also it is clear that the steady state flow stresses match, at least within the fluctuations--this is expected since the isomorph was constructed to have invariant flow stress. Nevertheless it serves as a check that the analytic formula for constructing the isomorph is reliable. However, the peak in the stress-strain curve is clearly not invariant--it decreases systematically with increasing density, by about 20\%, as the density rises to 2.8. Thus we have a clear deviation from isomorph invariance when the transient response to shearing is considered.  We remind the reader that the real ({\it i.e.}~non-reduced) shear stress involves a factor of $\rho k_BT$ which changes by over two orders of magnitude over the density range studied here.

\begin{table}
    \centering
    \begin{tabular}{c|c|c|c}
     & $\tilde{\dot{\epsilon}}=10^{-3}$ & $\tilde{\dot{\epsilon}}=10^{-4}$ & $\tilde{\dot{\epsilon}}=10^{-5}$  \\
    \hline
    $R_{c,0}=10^{-5}$ & $0.204 \pm 0.011$ & $0.213 \pm 0.014$ & $0.114 \pm 0.009$\\
    $R_{c,0}=10^{-6}$ & $0.211 \pm 0.006$ & $0.243 \pm 0.008$ & $0.236 \pm 0.012$\\
    $R_{c,0}=10^{-7}$ & $0.207 \pm 0.010$ & $0.242 \pm 0.006$ & $0.236 \pm 0.008$\\
    \end{tabular}
    \caption{Fractional change of $\tilde{\sigma _{p}}$ between the highest- and lowest-density isomorph points for different $R_{c,0}$ and $\tilde{\dot{\epsilon}}$ combinations.}
    \label{tab:peakstresschange}
\end{table}

\begin{table}
    \centering
    \begin{tabular}{c|c|c|c}
    & $\tilde{\dot{\epsilon}}=10^{-3}$ & $\tilde{\dot{\epsilon}}=10^{-4}$ & $\tilde{\dot{\epsilon}}=10^{-5}$  \\
    \hline
    $R_{c,0}=10^{-5}$ & 1.87 & 1.67 & 1.31\\
    $R_{c,0}=10^{-6}$ & 2.16 & 1.99 & 1.71\\
    $R_{c,0}=10^{-7}$ & 2.43 & 2.37 &  2.04\\
    \end{tabular}
    \caption{Ratio of $\tilde{\sigma _{p}}$ to $\tilde{\sigma _{f}}$ for the lowest-density isomorph points for different $R_{c,0}$ and $\tilde{\dot{\epsilon}}$ combinations.}
    \label{tab:peak_flow_ratio}
\end{table}

We find similar results for the other strain rates and cooling rates (details given in Appendix \ref{sec:AllSScurves}). Rather than show all of those stress-strain curves here, we instead extract the flow stress and peak stress from each stress strain curve, and plot these as a function of density in Fig.~\ref{fig:flow_sts_along_isomorph}(a) and (b), respectively; the curves themselves can be found in Appendix~\ref{sec:AllSScurves}.  Before considering the isomorphic behavior, we note that as expected, slower strain rates decrease both the peak stress and the flow stress; and slower cooling rates increase the peak stress but leave the flow stress unchanged.  The isomorphic behavior is evaluated through the dependence (or non-dependence) of these quantities on density along the isomorph.  In part (a) of the figure we see that the flow stresses are indeed flat within errors, as they have been constructed to be (the errors are comparable to, though smaller than, the symbol sizes). This plot also confirms our hypothesis that the isomorph determined by requiring invariant flow stress at one cooling rate and reduced strain rate is valid also for the others.
  
Part (b) of the figure shows the evolution of the reduced peak stress as a function of density along the isomorph. The trend is similar for all cooling and strain rates, with more or less similar relative drops of peak stress as density increases. In all cases the bulk of the drop occurs over densities $\rho_1\simeq1.3$ to $\rho_5\simeq 1.9$, after which the change in reduced peak stress for each 10\% increase in density is reduced. Interestingly the change between densities $\rho_0$ and $\rho_1$ is also smaller. Above density 1.9, $\tilde\sigma_p$ apparently continues to decrease linearly, and does not seem to have levelled off even at our largest density, though this must happen eventually as the Lennard-Jones potential becomes dominated by the repulsive IPL-term.

To summarize the influence of density on reduced peak stress, Table~\ref{tab:peakstresschange} gives the magnitudes of the relative stress drops over the full density range for all strain and cooling rates. The only apparent trend here is that both faster cooling and faster shearing tend to give slightly smaller drops, around 20\% instead of around 24\%. An apparent outlier is the value for the highest cooling rate and the lowest strain rate (black triangles), where the change in reduced peak stress is only 11\%. This case corresponds to the least stable glass being very slowly deformed, and has the lowest peak stress to start with.  To provide a different view of the influence of cooling rate and strain rate, Table~\ref{tab:peak_flow_ratio} shows the ratio between peak stress and flow stress at the lowest density. This indicates indeed that the same case of lowest strain rate and fastest cooling has the lowest ratio of peak to flow stress at the lowest density, 1.31. However there does not seem to be, upon comparing Tables~\ref{tab:peakstresschange} and \ref{tab:peak_flow_ratio}, a general correlation between fractional drop of peak stress with increasing density, and initial ratio of peak to flow stress. The most that can be said probably is that when the latter ratio is very low, there is less contrast between the non-flowing and flowing states, in the sense that the microscopic barriers to be crossed are not much different to start with and therefore there is less room for variation along the isomorph.
  
When quantifying the observed deviations in peak stress, the crucial question is whether they are sufficiently large to warrant declaring them a breakdown of the isomorph theory, or sufficiently small to be able to say that approximate isomorph invariance is a good approximation for the observed behavior. We can compare the deviations along an isomorph to the differences visible in Fig.~\ref{fig:firstpoint_crsr} due to variations in cooling and strain rates. In particular, the variation in peak stress along an isomorph is comparable to the difference associated with an order of magnitude change in cooling rate, as can be seen in Figs.~\ref{fig:stress_strain_along_isomorph_sr-3}-\ref{fig:stress_strain_along_isomorph_sr-5}. In this sense the variations are not ``small" and therefore we can speak of a breakdown of approximate isomorph invariance.
  
\subsection{\label{sec:equivalent_configs}Equivalent configurations in stress peak?}

\begin{figure}[htp]
    \centering
    \includegraphics[width=0.48\textwidth]{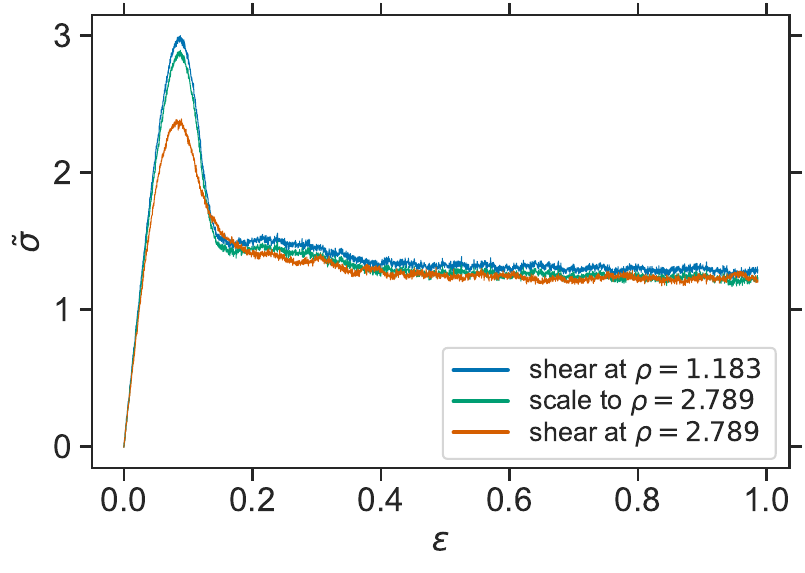}
    \caption{Comparison between stress-strain curves obtained from shearing at the lowest density (blue), computing the stress from the same configurations scaled to the highest density (green), and actually shearing at the highest density (orange) with $\dot{\tilde{\epsilon}}=10^{-3}$ and $R_{c,0}=10^{-7}$. }
    \label{fig:2ss}
\end{figure}

In attempting to understand the failure of the peak stresses to collapse, an important question is whether the configurations
sampled near the stress peak at different densities are equivalent. Equivalent means (statistically) indistinguishable after scaling to match densities. The simplest way to answer this question is to take configurations from near the peak in a simulation at one density, scale them to a different density and calculate the shear stress at the new density, thus generating a fictional stress-strain curve based on scaling configurations statically. This is similar to what is done in the stress-DIC method proposed above, but rather than use it to generate an isomorph we use it to compare the potential energy surface sampled by the same reduced configurations at different densities. If the ``fake" high-density stress-strain curve matches that actually simulated at high density then the conclusion would be that configurations at high stress are essentially equivalent to the corresponding ones at low densities but that the interactions are softened more at high densities and high stress than for high densities at lower stress (i.e. the steady state). If they do not match then something different must be happening in the microscopic dynamics during the stress peak. Fig.~\ref{fig:2ss} shows the result of this check. The curve generated from configurations sampled at the reference density $\rho_0$ and scaled $\rho_9$ (green) matches the curve at the reference density well, except for a small difference in normalization, and does not match the curve obtained from simulating at the high density $\rho_9$, which has the lower stress peak. This shows that the second possibility mentioned above must be the case: the particles undergo non-equivalent motion when simulated at the higher density. We examine what this non-equivalence is in the next section.

\section{Microscopic behavior}

The results of the previous section concern macroscopic mechanical properties. In this section we study variation of microscopic struture and dynamics, both at the particle level and at slightly larger length scales where we study inhomogeneities in the strain profile. The results of these different analyses suggest an explanation for the failure of the stress peak to collapse.

  \subsection{\label{sec:invariance_struc_dynamics}Invariance of pair structure and single-particle dynamics in the steady state}
  
\begin{figure}[htp]
    \centering
    \includegraphics[width=0.48\textwidth]{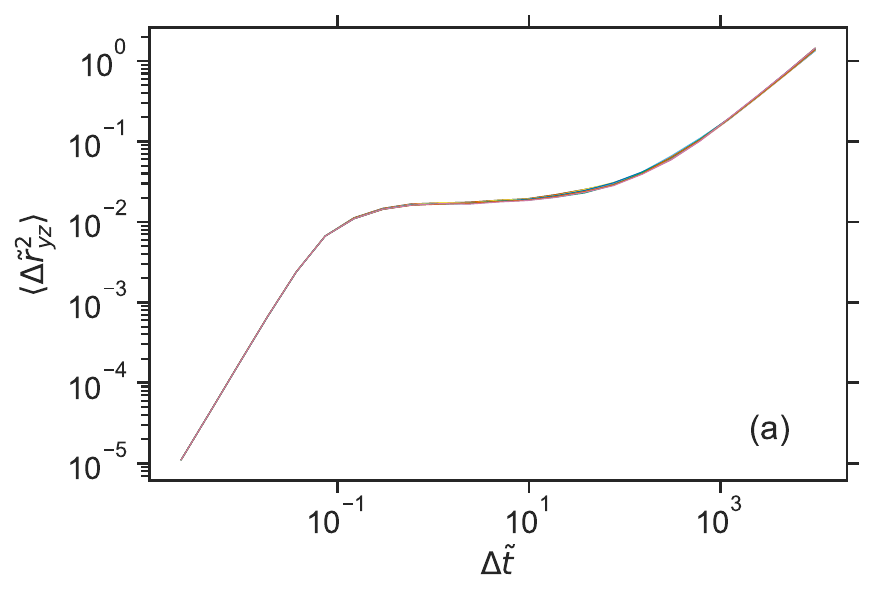}
    \includegraphics[width=0.48\textwidth]{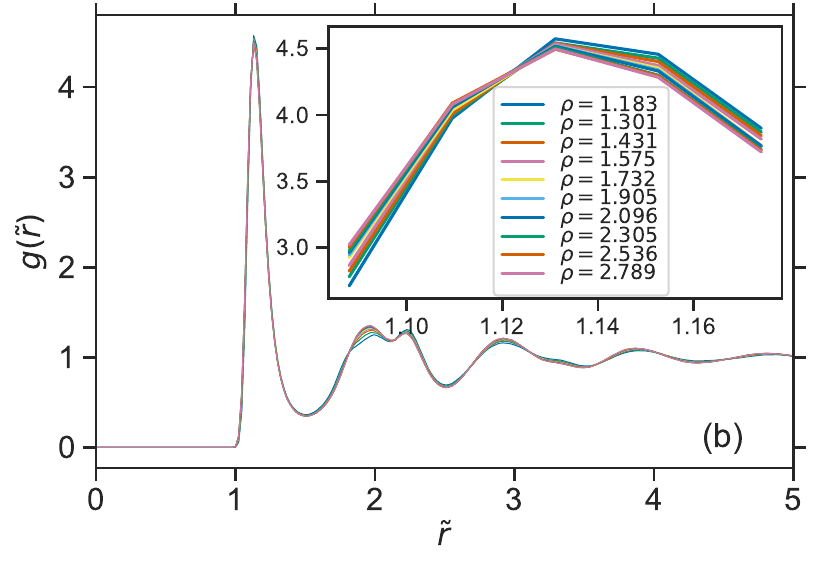}
    \caption{Collapse of (a) $\langle \Delta \tilde{r}_{yz}^{2} (\tilde{t}) \rangle$ and (b) $g(\tilde{r})$ (for AA pairs) for steady state only along an isomorph containing 10 points in the phase diagram calculated using the method described in subsection~\ref{sec:matching_flow_stress}. The starting configuration is at $\rho_0=1.183$ and $T_0 = 0.3$ and was cooled with $R_{c,0}=10^{-5}$. Shearing for all state points was with reduced strain rate $\dot{\tilde{\epsilon}}=10^{-4}$. The inset in (b) shows a close-up of the first peak.}
    \label{fig:msd}
\end{figure}
  
  In this subsection we consider particle-level measures of structure and dynamics properties, in particular self-diffusion and pair-structure. To investigate self-diffusion we plot in Fig.~\ref{fig:msd}(a) the mean squared transverse displacement (MSD) in reduced units, as a function of reduced time, for all densities along the isomorph at reduced shear rate $10^{-4}$. By transverse we mean that only components of displacement orthogonal to the shearing direction are included. The quality of the collapse is extremely good, so that it is not obvious to the eye that there are  in fact ten curves plotted. This plot is based on data from a single run for each density, since there is sufficient averaging over particles to get good statistics for single-particle dynamics.
  
  Fig.~\ref{fig:msd}(b) shows the radial distribution function (RDF) for AA pairs along the same isomorph. The collapse here is also very good, similar to what is seen in equilibrium liquids, including the slight deviations in the second peak\cite{Gnan2009}. The collapse for AB pairs (see Fig.~\ref{fig:gr_abbb} in the appendix) is poorer, with about a 15\% decrease in first peak height with increasing density, but it is still consistent with previous results, given the large density change involved. Both the MSD and RDF are determined from configurations drawn from the steady state. The quality of the invariance apparent in Fig.~\ref{fig:msd} confirms that the isomorph has been well determined. The non-invariance evidenced by the failure of the peak stress to collapse seems therefore to be restricted to the transition from non-flowing to flowing states. This will be discussed more below.

\subsection{\label{sec:spatial_variation}Variation in spatial homogeneity}

We now present results of analyzing strain profiles and local measures of plastic activity in order to determine whether the variation of peak stress can be associated with some systematic difference in the spatial organization of the initiation of flow, for example if flow is more or less localized in the peak at higher densities compared to lower densities.

\begin{figure}[htp]
    \centering
    \includegraphics[width=0.48\textwidth]{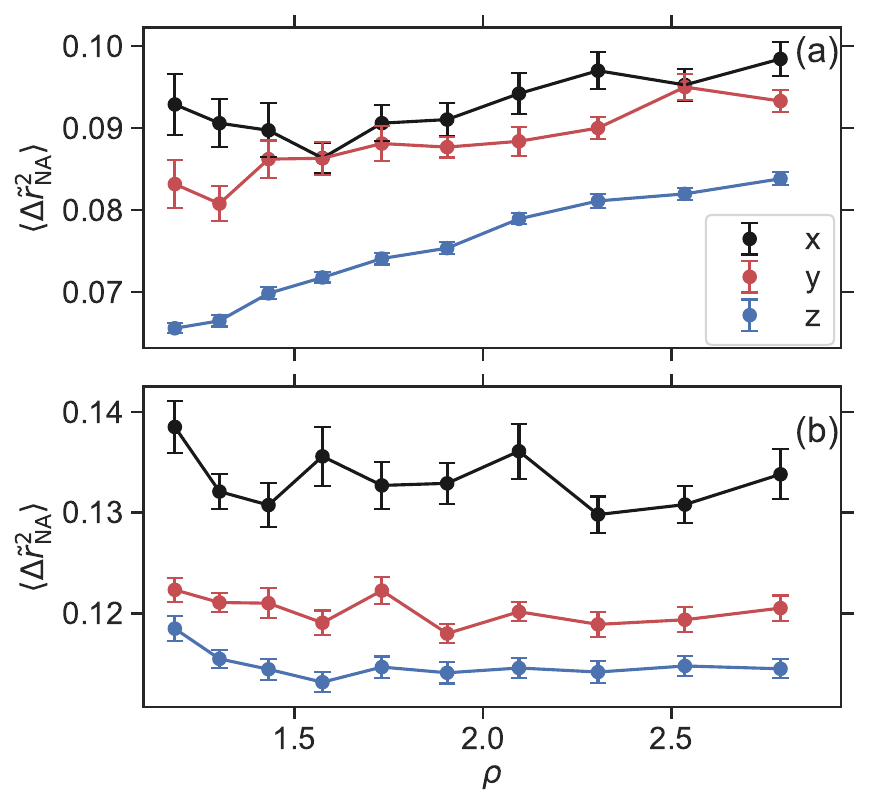}
    \caption{Comparison of non-affine particle motion over 12\%-strain intervals between (a) transient state with $\epsilon \in [0,0.12]$ and (b) steady state with $\epsilon \in [3.65,3.77]$. Data is averaged over all particles and 40 independent runs. Color indicates the three components of the nonaffine motion squared and shown in the legend in (a).}
    \label{fig:ndr2}
\end{figure}

The first indicator we consider is the non-affine displacement at the particle level. Recall that shearing motion occurs in the $x$-direction, while the gradient is in the $y$-direction. We first define the affine displacement $\Delta \vec{r}_{A,i}$ for particle $i$ as $\Delta \vec{r}_{\rm A,i}= \Delta \epsilon \,y_{i} \hat{x}$, where $y_{i}$ is the position of particle $i$ in the gradient direction and $\hat{x}$ is the velocity direction. This is simply the displacement associated with the macroscopic strain imposed on the system, which is known. Under shear, local rearrangements cause deviations from this affine motion. We define such deviation as the non-affine motion $\Delta \tilde{\vec{r}}_{\rm NA,i} = \Delta \tilde{\vec{r}}_{\rm real,i} - \Delta \vec{r}_{\rm A,i}$, where $\Delta \tilde{\vec{r}}_{\rm real,i}$ is the full displacement of particle $i$ \cite{Langer/Liu:1997,Goldenberg/Tanguy/Barrat:2007,Chen2010}. We expect that each component of the non-affine displacement is symmetrically distributed about zero, but it is possible that the distributions for the different components could be different. Figure~\ref{fig:ndr2} shows the variance of the non-affine displacement for each component, in reduced units. The data in part (a) of the figure are for displacements between undeformed configurations (strain 0) and corresponding configurations deformed to strain 0.12. The figure thus contains information about particle motions during the transition from the non-flowing to the flowing state, specifically those associated with the stress peak. As a function of increasing density the variance for each component increases, indicating a non-isomorph invariant behavior. In part (b) of the figure, data for a similar strain interval, but taken from the steady state regime is shown. Here little systematic variation as a function of density is apparent. Systematic differences between components are visible, with the variance for the $x$-component of non-affine displacement being highest both in the transient case and in the steady state. Chen {\em et al.} studied distributions of non-affine motion in experiments on colloids\cite{Chen2010}; their Fig. 8 shows PDFs of non-affine motion in the three different directions. The distribution for the out-of-plane direction (their $y$-direction, corresponding to our $z$-direction) is slightly narrower than for the other two directions, which is consistent with what we see.

\begin{figure}
    \centering
    \includegraphics[width=0.48\textwidth]{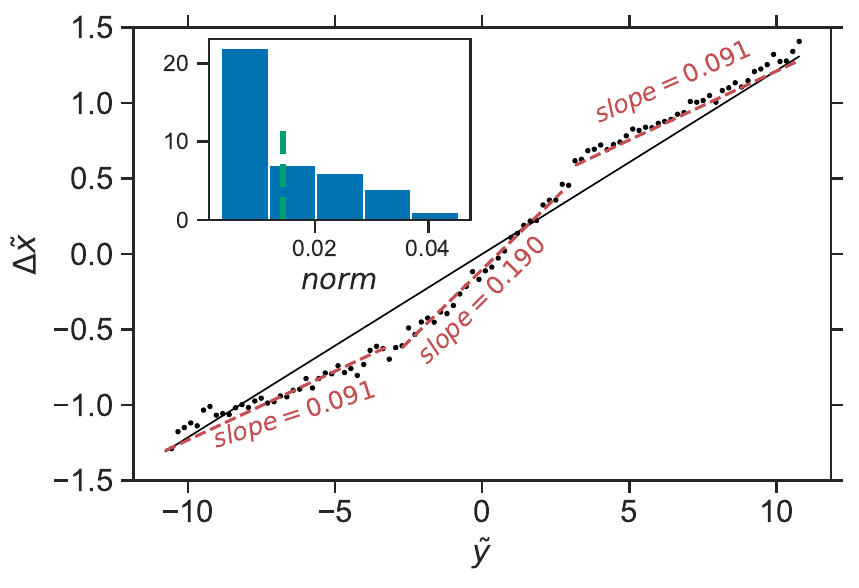}
    \caption{Example of a displacement profile exhibiting shear-banding. Displacement is calculated in steady state between $\epsilon=3.65$ and $\epsilon=3.77$ from a shear simulation at $\dot{\tilde{\epsilon}}=10^{-4}$ from a configuration at $\rho=1.301$ originally cooled at $R_{c,0}=10^{-5}$, by averaging the $x$-component of the particles' displacements over all particles within a bin defined by their $y$-coordinate. The straight line is the affine displacement following the applied strain. Dotted red lines are linear fits at different regions with two distinct slopes (treating the leftmost and rightmost regions together due to periodic boundary conditions). The norm, defined as the root mean square deviation of the bin values from the linear profile, is 0.022. The inset shows the histogram of norms of all 40 configurations at the corresponding density, cooling rate, and shear rate. The vertical dashed line indicates the mean value of the norm.}
    \label{fig:displacement_profile}
\end{figure}

\begin{figure}
    \centering
    \includegraphics[width=0.48\textwidth]{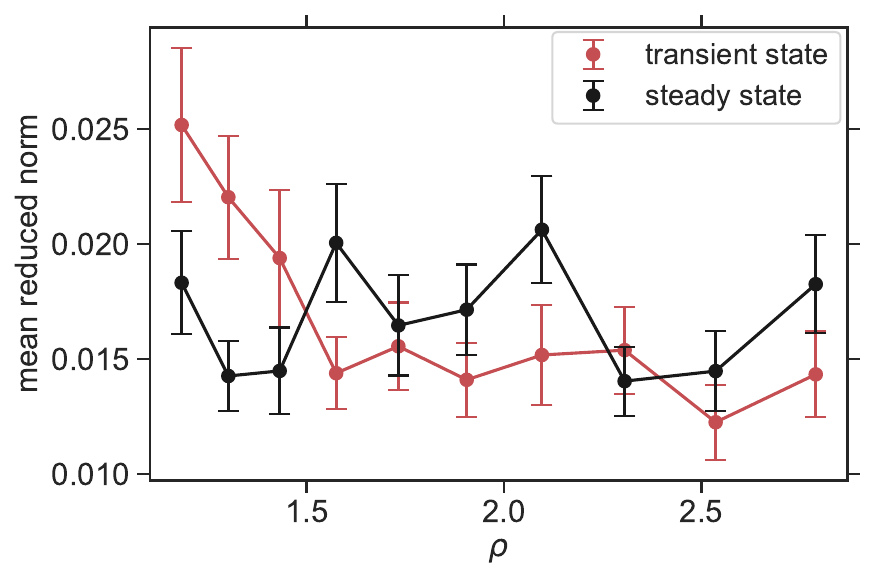}
    \caption{Comparison of norm of non-affine motion in velocity direction averaged of 40 configurations over 12\%-strain intervals between transient state with $\epsilon \in [0,0.12]$ and steady state with $\epsilon \in [3.65,3.77]$. Errors are the standard deviation of 40 divided by square root of 40. The definition of norm is the mean squared deviation of the displacement profile in Fig. \ref{fig:displacement_profile}.}
    \label{fig:norm}
\end{figure}

Another, related measure of spatial inhomogeneity is to consider the displacement profiles obtained by binning particles according to their $y$-coordinate and averaging the $x$-displacement for all particles in a bin. This gives a probe of systematic variation in the gradient direction, while averaging over other directions. An example is shown in Fig.~\ref{fig:displacement_profile} for a 12\% strain interval from the steady state in a particular run. A clear systematic deviation from the affine profile is visible: the system exhibits two distinct regions with the strain differing by a factor of two as indicated by the linear fits (note that the region of the left is connected to that on the right via the periodic boundary conditions). This coexistence of regions with differing strain (rates) is termed shear-banding\cite{Bailey/Schiotz/Jacobsen:2006,Ovarlez/Others:2009,Martin/Hu:2012,Greer/Cheng/Ma:2013}.

Profiles for different runs exhibit somewhat similar shapes, with mostly a single region of higher strain, more or less sharply delimited from the rest of the system, with varying contrast ({\it i.e.,} difference in strain). To quantify this contrast and look for systematic variations along the isomorph we take the mean squared deviation of each point from the affine line, yielding a norm of the (non-affine) displacement profile. For the example in Fig.~\ref{fig:displacement_profile} the norm is 0.022; a histogram of norms for that density is shown in the inset, and exhibits a somewhat broad distribution. Note this measure of non-affine motion differs from that presented in Fig.~\ref{fig:ndr2} in the initial averaging within a $y$-bin {\em before} squaring and further averaging over bins. We calculate the mean norm from the 40 runs at each density, and plot these as a function of density in Fig.~\ref{fig:norm}. As with Fig.~\ref{fig:ndr2} we see a clear trend in the transient data, and no discernible trend in the steady state data. This suggests that at lower density there is a greater tendency for quasi-shear-banding to occur during the stress peak. This decreases with increasing density, particularly over the first four densities. The steady-state values in panel (b) are mostly below the low-density transient values; there are a couple of exceptions to this but no overall trend. We have only looked at a specific interval covering 12\% of strain in all cases; an interesting question for future work is to what extent a given quasi-shear-band persists over longer amounts of strain. Interestingly, and apparently counterintuitively, the trend visible for the red data points in Fig.~\ref{fig:norm} is in the opposite direction to those shown in Fig.~\ref{fig:ndr2}(a). What these two results tell us is that there is a {\em increase} of ``incoherent non-affine motion" (i.e., measured at the particle level) as density increases, while there is a {\em decrease} of ``coherent non-affine motion" (i.e., the signal that remains after averaging over displacements within a horizontal slice).

\section{Discussion}

The first part of this work was concerned with identifying candidate isomorphs in non-equilibrium situations. The force method, and other methods inspired by those for determining isomorphs in equilibrium, generate candidate isomorphs along which the reduced-unit flow stress is reasonably constant, within a few percent, for moderate density changes, but can vary up to $\pm$ 20\% for extreme density changes (over a factor of two), corresponding to an uncertainty in the estimated isomorphic temperature of $\pm$ 10\%. The 20\% should be compared to an overall variation of over a factor of 100 in the real flow stress, however, and a factor of 50 in the overall temperature change. Clearly, obtaining a
precise estimate of isomorphic temperatures for such large density jumps is asking too much. Given these difficulties our strategy has been to require the reduced flow stresses to match at all densities as a practical method of identifying the isomorph. This can be considered somewhat analogous to choosing contours of the excess entropy in equilibrium, and has the advantage of being well-defined and straightforward to calculate in simulations of glasses, unlike Dyre's systemic temperature \cite{Dyre2020}, for example. In both cases a choice is made to identify isomorphs as contours of one particular quantity, while the rest of the work involves investigating the invariance of other quantities.

Because of intrinsic fluctuations in the shear stress during steady state deformation\cite{Jiang2019}, an accurate determination of the flow stress requires a combination of long runs and/or many independent runs, and is therefore computationally demanding. This is even more the case when one considers the extra work of doing so at several temperatures in order to identify the matching temperature. But by hypothesizing that this only needed to be done for one or two densities, other than the starting density, by applying the analytic formulas that describe the shape of isomorphs for Lennard-Jones systems, and moreover doing this only for one combination of strain and cooling rates, we save a lot of work. These assumptions were justified by the resulting flat plots of reduced flow stress as a function of density. Further evidence of the quality of our choice of determining isomorphic temperatures for given density jumps is given by the invariance of microscopic dynamics and structure, Fig.~\ref{fig:msd}.

The fact that the force and fluctuation-based methods diverge for large density jumps is in the end not surprising since isomorph invariance is only approximate for most potentials. Moreover, trying to understand the differences in terms of systematically different ways in which different parts of the potential energy surface -- for example those relevant for vibrational motion versus barrier-crossing flow events -- would probably not be fruitful. Some insight might be gained, however, by investigating the barriers to flow explicitly using barrier finding techniques on small systems to trace the way a typical energy barrier scales with increasing density.

The second main result of this work is the failure of the (reduced) peak stress under shear deformation to collapse along the isomorph determined by forcing collapse of the flow stress. Simply put, no temperature can match both peak and flow stress, meaning that the relevant energy barriers apparently scale differently with density. This is despite that not just the flow stress (by construction) but also the initial linear part of the stress-strain curve are invariant, along with the particle motions and most pair correlations during the steady state. The analysis of section~\ref{sec:equivalent_configs} shows that the variation in peak stress is not simply a question of the same trajectories (in reduced coordinates) experiencing different potential energies, forces and stresses, but rather the trajectories are non-equivalent in the region of the stress peak, as evidenced by the analysis of non-affine motion. For different (reduced-unit) trajectories to result, the (reduced-unit) forces have to be different, of course. But subtle differences in the forces can lead to macroscopic differences in the trajectories, in particular the degree of shear-banding, which then leads to pronounced differences in the observed stresses. A possible explanation for a variation in the tendency for shear-banding could be that it depends sensitively on the steepness (effective IPL exponent) of the potential, something that has apparently not been tested before in the literature. This could be tested by running IPL simulations with different exponents and studying the tendency to create shear bands. A connection between shear banding and non-invariance of the macroscopic stress strain curve would be analogous to the case of the melting curve, which is only approximately an isomorph\cite{Pedersen/others:2016a}. In that case the two co-existing phases, having different densities, cannot be expected to scale identically. Likewise here, one could the presence of co-existing shear bands is a plausible source of a breakdown in isomorph invariance for the whole system. It is not due to differing densities, though: We have not found a significant difference in density between the differently shearing coexisting regions.

The range of effective IPL exponents is quite limited for Lennard-Jones systems, varying from around 18 at low pressures to approaching 12 at the highest pressures. Potentials which exhibit more dramatic variation of effective exponent include the exponential pair potential \cite{Bacher/Schroder/Dyre:2018} and the many-body effective medium theory potential for metallic systems\cite{Friedeheim2019}. Studying these systems would give additional insight, by potentially exhibiting an even more pronounced variation in the peak stress along isomorphs; it might for example be noticeable at relatively small density changes.

A technical point should be raised here. Our approach using the SLLOD algorithm is predicated on an assumed linear strain profile, i.e. a single global strain rate, and is thus technically inconsistent with the occurrence in practice of a certain degree of shear-banding. This is in principle problematic because the velocities used to define the kinetic energy are defined with respect to the assumed linear streaming velocity profile. Therefore in principle some of the (fixed) total kinetic energy goes into the deviations of the streaming velocity from a linear profile\cite{Evans/Morriss:1986,Bagchi/Others:1996} and less is available for thermal motion. We note, however, that the typical deviation of the real streaming velocity from the linear profile is of order 0.1\% of the thermal velocity, so any effects on the effective temperature are negligible. \newpurple{To confirm that the observed behavior is not an artifact of the thermostat, we have carried out some simulations with a modified thermostat by which only the kinetic energy associated with velocities in the non-flow directions ($y$ and $z$) is thermostatted, while the part associated with velocities in the flow direction ($x$) is free to fluctuate. We find the same behavior, in particular the trends noticed in Fig. 12 are even clearer with the modified thermostat.}

Finally, we note that since Fig~\ref{fig:norm} indicates that shear banding occurs also in the steady state, a varying tendency towards shear banding depending on density (via the effective IPL exponent) could actually be present in the steady state. Since we have chosen the isomorph temperatures to match the reduced steady-state shear stress we do not see this in our data (essentially because temperatures have been adjusted to compensate for it), but it could potentially underlying the failure of the fluctuation methods to predict the steady-state stress. Comparing different IPL systems under steady state shear could also shed light on this.

\begin{acknowledgments}

This material is based upon work supported by the
National Science Foundation under Grant No. (CBET-
1804186) (YJ and ERW) and the VILLUM Foundations's {\em Matter} Grant (No. 16515).  The data will be made publicly available via Emory University's Dataverse upon publication.

\end{acknowledgments}

\appendix

\section{\label{rheology}Effects of strain rate and cooling rate}

\begin{figure}[htp]
  \includegraphics[width=0.48\textwidth]{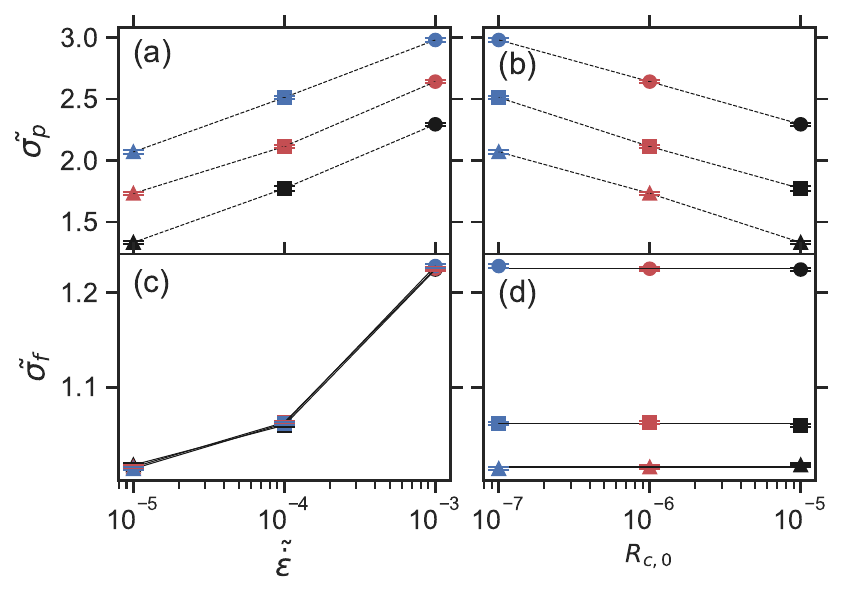}

  \caption{\label{fig:firstpoint_crsr} Reduced peak stress $\tilde{\sigma}_{p}$ and flow stress $\tilde{\sigma}_{f}$ as function of reduced strain rate $\tilde{\dot{\epsilon}}$ (a,c) and cooling rate $R_{c,0}$ (b,d). Black, red, and blue are for $R_{c,0} = 10^{-5}, 10^{-6}, 10^{-7}$; sphere, square, and up triangle are for $\tilde{\dot{\epsilon}} = 10^{-3}, 10^{-4}, 10^{-5}$. The flow stress $\tilde{\sigma}_{f}$ is within errors independent of $R_{c,0}$ but increases with $\tilde{\dot{\epsilon}}$.  See the text for a discussion of the uncertainties of these measurements, which for these data are smaller than the symbol size.  The three horizontal lines in panel (d) are the average of the corresponding three points of the same $\tilde{\dot{\epsilon}}$.}
\end{figure}

In this appendix we present an overview of the basic rheological properties of our system; specifically we show flow stress and peak stress for different strain rates and cooling rates. We restrict attention to a single density-temperature state point, namely $\rho_0=1.183,T=0.3$ which in the main text is our initial, or reference state point for constructing isomorphs.

\begin{figure}[htp]
  \includegraphics[width=0.45\textwidth]{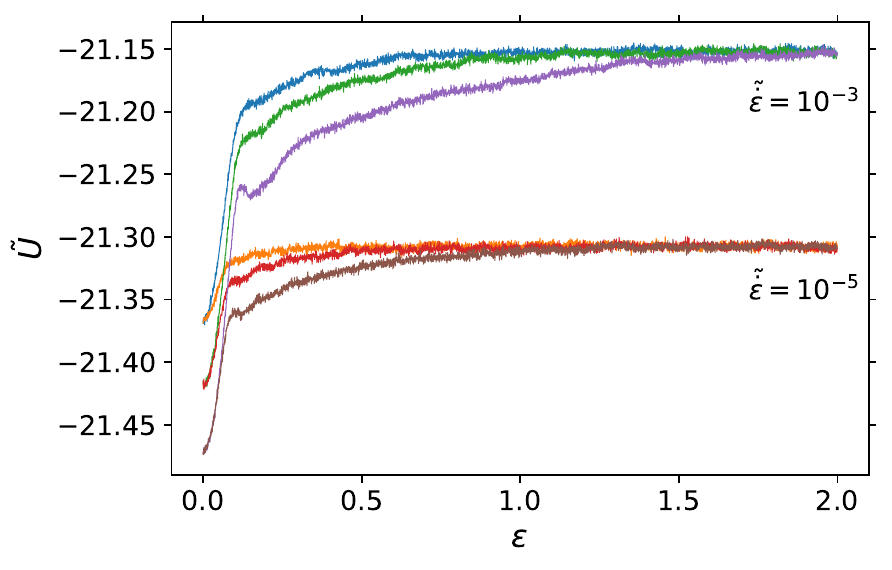}
  \caption{Reduced potential energy against strain sheared with $\tilde{\dot{\epsilon}} = 10^{-3}, 10^{-5}$ for the three cooling rates $R_{c,0} = 10^{-5}$ (blue and orange), $10^{-6}$ (green and red), $10^{-7}$ (purple and brown) at $\rho = 1.183$ and $T=0.3$. Each curve is an average of 40 simulations.}
\label{fig:pe_curves}
\end{figure}

Since we do not expect perfect isomorphs, both because isomorphs are never perfect, especially over large density changes, and because the intrinsic fluctuations in the stress can obscure the degree of collapse, we need to be able to compare potential ``approximate invariance'' with the variation observed when non-isomorphic changes of parameters are considered. In equilibrium situations one can, for example, vary the temperature and density separately while keeping the other fixed. In the non-equilibrium case more interesting possibilities arise, namely varying the cooling and strain rates. To show how much variation in rheological properties results from varying these rates we plot the peak and flow stresses in Fig.~\ref{fig:firstpoint_crsr}, first as a function of strain rate for different cooling rates, and then as a function of cooling rate for different strain rates. It can be seen that an order of magnitude increase in strain rate increases the peak stress by 25-30\%  [panel (a)] and the flow stress by 5-15\% [panel (c)] while an order of magnitude increase in cooling rate decreases the peak stress also 15-20\% [panel (b)] but has no effect on the flow stress [panel (d)]. These dependencies are expected in glassy rheology\cite{Voigtmann:2014}. In particular a lower cooling rate generates a more stable glass, which requires a larger stress to initiate deformation (i.e. it has a larger yield stress).

Fig.~\ref{fig:pe_curves} shows the potential energy versus strain up to strain 2 for the three cooling rates and the fastest and slowest strain rates. The potential energy can be rather slow in converging to its steady state value, especially for the highest strain rate and lowest cooling rate, where it appears to converge around strain 2 (we can be sure the purple curve has converged since it must converge to the same value as the other curves for the same strain rate). This is despite that the shear stress has typically converged before strain 1. The fact that potential energy must also be monitored to ensure steady state conditions was discussed by Singh {\em et al.} \cite{Singh2020}.

\section{\label{sec:force-method-systemic-temperature}Connecting the force method to the systemic temperature}

We start with the definition of systemic temperature, $T_s$, from  Dyre \cite{Dyre:2018a,Dyre2020}, where the function $U(\rho, \Sex)$ should, initially, be taken as the equilibrium thermodynamic relation between potential energy, density and excess entropy:

\begin{equation}
T_s(\bfa{R}) \equiv \frac{\partial U(\rho, \Sex(\bfa{R}))}{\partial \Sex}
\end{equation}
Here the microscopic definition of the excess entropy $\Sex(\bfa{R})$ for individual configurations is also based on the same equilibrium relation between the thermodynamic quantities, and the potential energy of a given configuration (it is the excess entropy for the thermodynamic state whose mean potential energy is equal to the configuration's potential energy).

To get a different expression for $T_s$, consider the derivative of $U$ with respect to $\tilde {\bfa{R}}$, at fixed $\rho$.  If we consider a small change, at fixed density, of one of the reduced reduced coordinates, say atom $i$, spatial coordinate $\alpha$, then by the chain rule, we have

\begin{equation}\label{dUdR_chain_rule}
\frac{\partial U(\rho, \Sex)}{\partial \tilde R_{i,\alpha}} = \frac{\partial U(\rho, \Sex)}{\partial \Sex}
  \frac{\partial \Sex(\rho, \tilde{\bfa{R}} )}{\partial \tilde R_{i,\alpha}}
\end{equation}
The first factor on the right is just $T_s$; but before we solve for it we wish to involve all coordinates. Noting that $T_s$ is the same independent of which coordinate we choose, we square Eq.~\eqref{dUdR_chain_rule}, sum over $i$ and $\alpha$, and finally take the square root, giving


\begin{equation}\label{dUdR_chain_rule_sum_sq}
\sqrt{\sum_{i,\alpha}\left|\frac{\partial U(\rho, \Sex)}{\partial \tilde R_{i,\alpha}}\right|^2} = T_s \sqrt{\sum_{i,\alpha} \left|\frac{\partial \Sex(\rho, \tilde{\bfa{R}} )}{\partial \tilde R_{i,\alpha}} \right|^2}
\end{equation}
So our new expression for $T_s$ is 
\begin{equation}
  T_s = \frac{\sqrt{\sum_{i,\alpha}\left|\frac{\partial U(\rho, \Sex)}{\partial \tilde R_{i,\alpha}}\right|^2} }
  {\sqrt{\sum_{i,\alpha} \left|\frac{\partial \Sex(\rho, \tilde{\bfa{R}} )}{\partial \tilde{R}_{i,\alpha}} \right|^2}}.
\end{equation}
This is still an exact expression. Moreover the dependence of $U$ on $\Sex$ can be dropped by reinterpreting the function $U$ as the microscopic potential energy: thus, the numerator can be written as $\rho^{-1/3}|\bfa{F}|$ where $\bfa{F}$ is the 3N-dimensional force vector for the whole system. If we now consider the ratio of $T_s$ at two different densities, $\rho_1$ and $\rho_2$, for the same reduced coordinates, we have

\begin{equation}\
  \frac{T_s^{(2)}}{T_s^{(1)}} = \left[ \left(\frac{\rho_1}{\rho_2}\right)^{1/3} \frac{|\bfa{F}_2|}{|\bfa{F}_1|} \right]
\frac
{\sqrt{\sum_{i,\alpha} \left| 
\frac{\partial \Sex(\rho_1, \tilde{\bfa{R}})}{\partial \tilde{R}_{i,\alpha}} 
\right|^2}
}
{{\sqrt{\sum_{i,\alpha} \left|\frac{\partial \Sex(\rho_2, \tilde{\bfa{R}} )}
{\partial \tilde{R}_{i,\alpha}} \right|^2}}
}
\end{equation}

The first factor (in square brackets) on the right side is exactly the expression for the temperature ratio given by Schr{\o}der's force method\cite{Schroder:2022}. What about the second factor, involving the derivatives of $\Sex$, with respect to reduced coordinates, at different densities? For perfect hidden scale invariance where $\Sex$ does not depend on $\rho$, this factor is unity, and in that case, not surprisingly, the ratio of systemic temperatures is the ratio given by the force method (or any other method). In the case of imperfect scaling, we can argue that this factor is nearly unity: we take a derivative with respect to $\tilde R_{i,\alpha}$, so even if changes in $\Sex$ accumulate over large density changes, these changes will almost all cancel when we compare two nearby values of $\tilde R_{i,\alpha}$. Thus the derivatives depend much less on density, and we can assume that this factor is well approximated by unity, giving the desired result.


\section{Methods for identifying isomorphs in equilibrium}
\label{subsec:methods}

There are several methods currently in use for identifying isomorphs in equilibrium. We review these methods here for two reasons.  Firstly because we attempt to extend these methods to the out-of-equilibrium systems of interest (discussed in Sec.~\ref{sec:identify_iso_noneq}).  Secondly, because our final procedure involved the use the analytic method (Sec.~\ref{subsec:analytic}) to interpolate our isomorph curve in our out-of-equilibrium system. 

\subsection{Integration using the density scaling exponent $\gamma$}
In equilibrium the slope of isomorphs in the $\ln\rho, \ln T$ plane is given by the so-called density scaling exponent  $\gamma$, defined generally as the slope of configurational adiabats--curves along which the excess entropy is constant\cite{Gnan2009}:

\begin{equation}\label{eq:gamma_definition}
    \left(\frac{\partial T}{\partial\rho}\right)_{\Sex} = \gamma (\rho, T) = \frac{\langle\Delta U\Delta W\rangle}{\langle(\Delta U)^2\rangle}
\end{equation}
where the second equality indicates how $\gamma$ is determined from fluctuations at a particular state point. Angle brackets represent NVT ensemble averages, and the last expression is simply the linear regression slope of a scatter plot of $W$ against $U$.
The virial $W$ can be defined as the derivative of $U$ for a configuration with respect to $\ln\rho$, where $\rho\equiv N/V$ is the number density of the system. When taking the derivative it should be understood that the number and relative positions of the particles are kept fixed and only a uniform scaling is involved. Thus $W$ contains information about how the potential energy surface changes under (infinitesimal) uniform scaling and therefore is naturally relevant for the identification of isomorphs. From the same linear regression fit a correlation coefficient $R$ may be extracted, which is used to gauge the expected quality of the isomorphs. By determining $\gamma$ from fluctuations an isomorph in equilibrium may be traced by simple numerical integration (explicit Euler method) of Eq.~\eqref{eq:gamma_definition}, taking small steps in density (typically 1\%, although larger jumps are possible with higher order integration techniques\cite{Attia/Dyre/Pedersen:2021}). For systems with interactions described by an inverse power law (IPL) with a particular exponent $n$, exact isomorphs exist and the density scaling exponent is $n/3$; otherwise it depends mainly on density, though it does have some temperature dependence\cite{Bailey/others:2008b}.

\subsection{Direct isomorph check}

An early formulation of isomorphism involves the proportionality of Boltzmann factors of corresponding microscopic states. Here ``corresponding" means all particles being the same in reduced coordinates, i.e., one configuration is obtained from the other by a uniform scaling. A consequence of this proportionality, obtained simply by taking logarithms, is a proportionality between scaled and unscaled potential energies

\begin{equation}\label{eq:DIC}
    U_f(\bfa{R_f}) = \frac{T_f}{T_i} U_i(\bfa{R_i}) + \textrm{const.},
\end{equation}
where subscripts $i$ and $f$ indicate potential energies evaluated at initial and final densities, respectively for any configuration with given reduced coordinates. Here we use the upper case boldface $\bfa{R}$ to represent the entire 3$N$-vector of particle coordinates, for convenience, and the equality of scaled coordinates can be expressed as $\bfa{\tilde R_f}\equiv\rho_f^{1/3}\bfa{R_f} = \rho_i^{1/3}\bfa{R_i}\equiv\bfa{\tilde R_i}$. Eq.~\eqref{eq:DIC} was originally considered a simple check of the basic isomorph concept of proportional energy fluctuations, hence the name direct isomorph check (DIC)\cite{Gnan2009}, but it also suggests a method for identifying isomorphs: Given densities $\rho_i$ and $\rho_f$, and the temperature $T_i$, the temperature $T_f$ such that state point $\rho_f,T_f$ is isomorphic to state point $\rho_i, T_i$ may be identified by (1) sampling configurations from an equilibrium simulation at $\rho_i, T_i$, (2) scaling them to density $\rho_f$, (3) calculating the potential energies of the scaled configurations, and (4) making a scatter plot of the scaled versus unscaled potential energies. Without requiring proportionality of Boltzmann factors one can also derive the DIC by considering configurational adiabats\cite{Schroder2014}. Furthermore, when considering infinitesimal density changes the DIC reduces to the method of integrating Eq.~\eqref{eq:gamma_definition}.

\subsection{Stress-based direct isomorph check}

One can interpret the DIC as choosing the temperature $T_2$ by requiring the reduced-unit energy fluctuations to be as close as possible between the two state points, where ``as close as possible" involves a linear regression fit. One can in principle make a similar requirement for other quantities, for example the virial, whose fluctuations should also be related by being the same in reduced units. Or indeed the shear stress (configurational part). The latter leads to an alternative version of the direct isomorph check where the (configurational part of) the shear stress for scaled configurations is plotted against that for the unscaled ones. In this case, in view of Eq.~\eqref{eq:sigma_tilde}, the slope of the linear regression should be $\rho_f T_f/\rho_i T_i$. This suggests an alternative method for identifying an isomorphic temperature which may be relevant in deformation simulations. The shear stress given by our code includes the (small) kinetic part by default; we have checked in one case that its presence make a negligible difference to the fitted slope.

\subsection{Analytic isomorph formula for LJ potentials}
\label{subsec:analytic}

For pair potentials an analytic formula describing the shapes of isomorphs is available\cite{Ingebrigtsen2012,Bohling2014}, which for the Lennard-Jones potential takes the form $T(\rho)\propto h(\rho)$, where the density scaling function $h(\rho)$ is given by

\begin{equation}
    h(\rho) = A\rho^4 - B\rho^2
\end{equation}
The analytic form of $h(\rho)$ is directly related to that of the potential (indeed it is essentially the second derivative of the pair potential, evaluated at $r=\rho^{-1/3}$ and expressed in reduced units)\cite{Bohling2014}. The overall normalization of $h(\rho)$ is undefined since there is a proportionality constant in the relation between it and the temperature, so there is in fact only one free parameter, which can be taken to be the ratio $B/A$. If this is known then given two densities, $\rho_i$ and $\rho_f$, and a temperature $T_i$ corresponding to density $\rho_i$, the temperature $T_f$ corresponding to density $\rho_f$ is given by

\begin{equation}\label{eq:isomorph_from_h_rho}
    T_f = T_i \frac{h(\rho_f)}{h(\rho_i)} = T_i \frac{\rho_f^4 - (B/A) \rho_f^2}{\rho_i^4 - (B/A) \rho_i^2}
\end{equation}
To fix the parameter $B/A$ two options are available. One can note that the logarithmic derivative of $h(\rho)$ must also be equal to the density scaling exponent $\gamma$,

\begin{equation}
    \gamma(\rho) = \frac{d\ln h(\rho)}{d\ln\rho} = \frac{4\rho^4 - 2 (B/A) \rho^2}{\rho^4 - (B/A) \rho^2},
\end{equation}
where we assume explicitly that $\gamma$ depends only on density. Considering a particular reference density $\rho_{ref}$ at which $\gamma$ is to be evaluated (for example by simulation), isolating $B/A$ gives

\begin{equation}\label{eq:B_A_from_gamma}
    \frac{B}{A} = \frac{(\gamma(\rho_{ref})-4)\rho_{ref}^2}{(\gamma(\rho_{ref})-2)}
\end{equation}
In principle, if isomorph theory was exact, one could run a single simulation at the reference density, evaluate $\gamma$ from the $U,W$ fluctuations there, use Eq.~\eqref{eq:B_A_from_gamma} to determine $B/A$ and generate the whole isomorph using Eq.~\eqref{eq:isomorph_from_h_rho}. This works reasonably well for small density jumps, say 10\% -- so certainly better than the integration method with steps of 1\% -- but it does not give accurate temperatures for very large density jumps. Rather, the greatest utility of the analytic isomorph expression is its use in {\em interpolating} between points known to be isomorphic to get the points between\cite{Bohling2012}. That is, if both densities and temperatures for two state points, $\rho_i, T_i$ and $\rho_f, T_f$ are known, Eq.~\eqref{eq:isomorph_from_h_rho} can be solved for $B/A$, giving

\begin{equation}\label{eq:B_A_from_two_points}
    \frac{B}{A} = \frac{\rho_f^4 - \rho_i^4 \frac{T_f}{T_i}}{\rho_f^2 - \rho_i^2 \frac{T_f}{T_i}}
\end{equation}
Then isomorphic temperatures for densities between $\rho_i$ and $\rho_f$ can readily be found using Eq.~\eqref{eq:isomorph_from_h_rho} (replacing $\rho_f$ and $T_f$ with the intermediate values). This is the manner in which we use the analytical formula in this work -- note that it does not explicitly depend on having equilibrium.

\section{\label{sec:AllSScurves}Stress strain curve collapse for other cooling and (reduced) strain rates}

\begin{figure}
\includegraphics[width=0.48\textwidth]{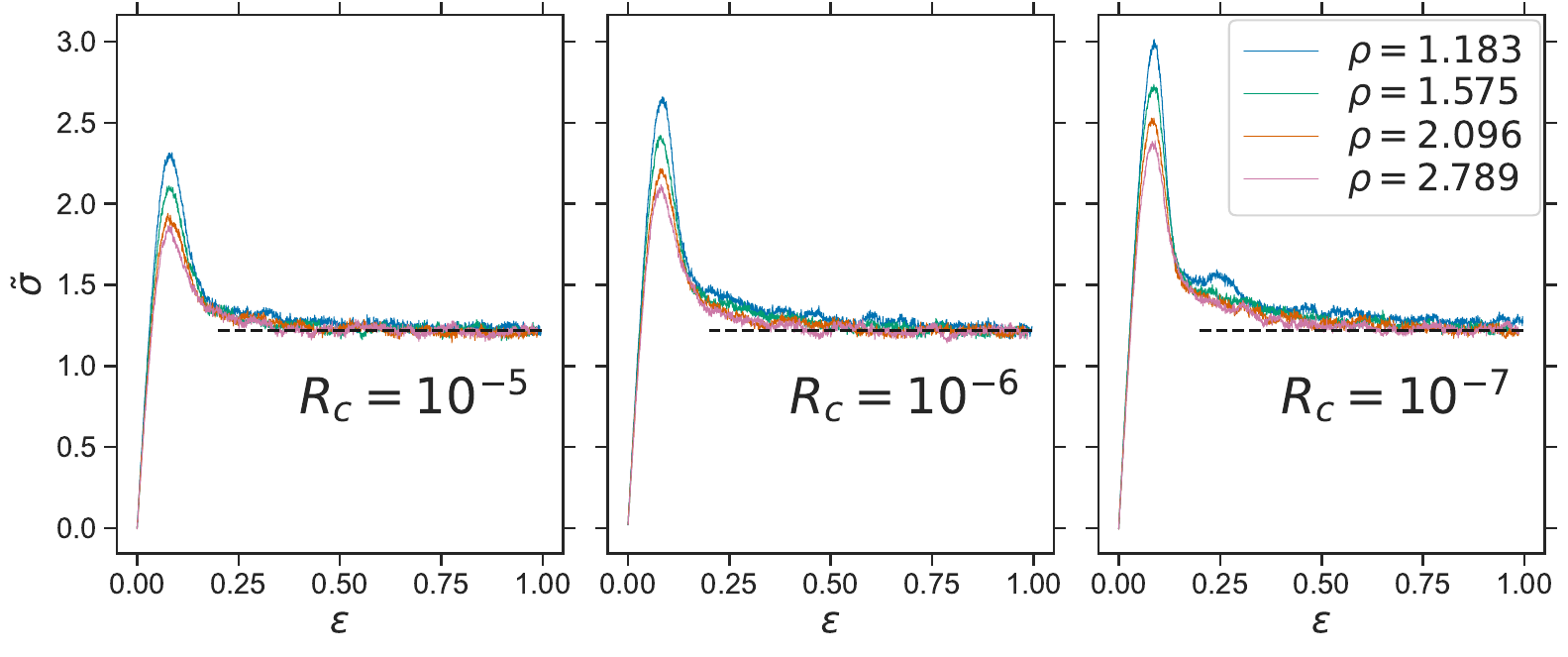}
\caption{\label{fig:stress_strain_along_isomorph_sr-3}Stress and strain curves of glasses cooled at three rates and sheared at reduced strain rate $\dot{\tilde{\epsilon}}=10^{-3}$. We only show strain up to 1 so that the peak is more visible; only $4$ densities are shown for clarity. The black horizontal dashed lines indicate the flow stress. These curves are where the data of Fig.~\ref{fig:flow_sts_along_isomorph} are derived from.}\end{figure}

\begin{figure}
\includegraphics[width=0.48\textwidth]{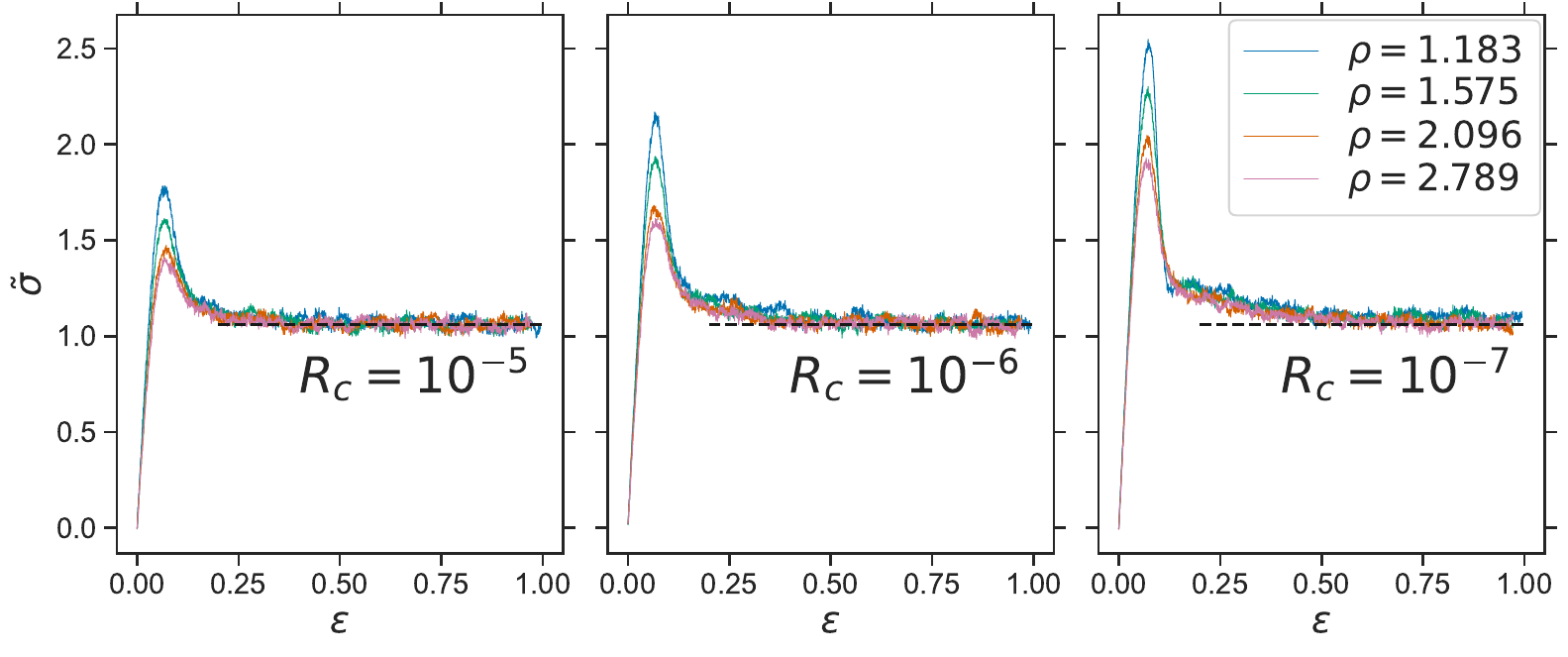}
\caption{\label{fig:stress_strain_along_isomorph_sr-4}Same as in Fig.~\ref{fig:stress_strain_along_isomorph_sr-3} but now for glasses cooled at three rates and sheared at reduced strain rate $\dot{\tilde{\epsilon}}=10^{-4}$.}
\end{figure}

\begin{figure}
\includegraphics[width=0.48\textwidth]{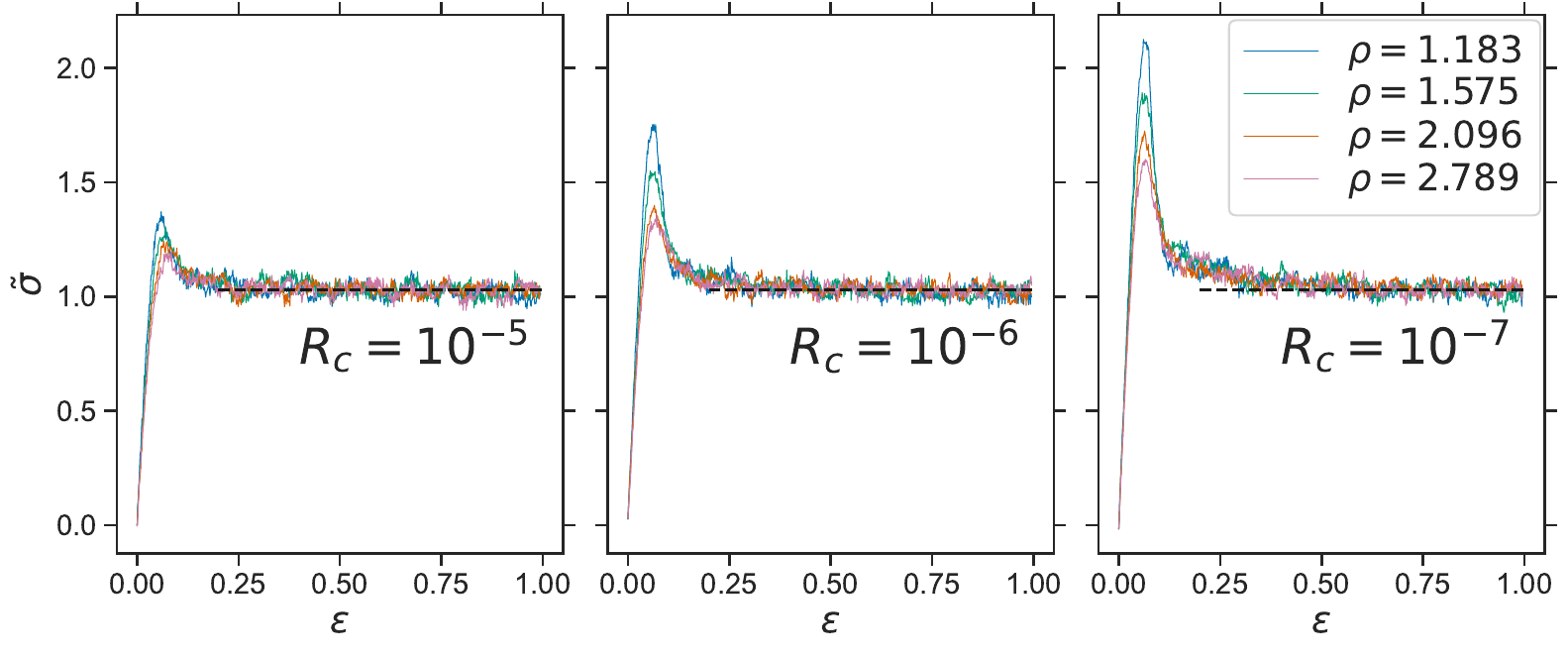}
\caption{\label{fig:stress_strain_along_isomorph_sr-5}Same as in Fig.~\ref{fig:stress_strain_along_isomorph_sr-3} but now for glasses cooled at three rates and sheared at reduced strain rate $\dot{\tilde{\epsilon}}=10^{-5}$.}
\end{figure}

We present here for completeness the full set of strain curves for all cooling and strain rates. Figs.~\ref{fig:stress_strain_along_isomorph_sr-3}, \ref{fig:stress_strain_along_isomorph_sr-4} and \ref{fig:stress_strain_along_isomorph_sr-5} show data for reduced strain rates $10^{-3}$, $10^{-4}$ and $10^{-5}$, respectively. The individual panels show data for the different cooling rates. By comparing panels within one figure one can see the effect of the changing cooling rate on the initial peak, while the effect of different strain rates can be seen in the flow stress--constant within each figure---varying from one figure to another. In all cases the behavior discussed in the main text, namely the decrease of reduced peak stress as a function of density along an isomorph, is clearly visible.

\section{Radial distribution functions for other types.}

\begin{figure}[htp]
  \includegraphics[width=0.45\textwidth]{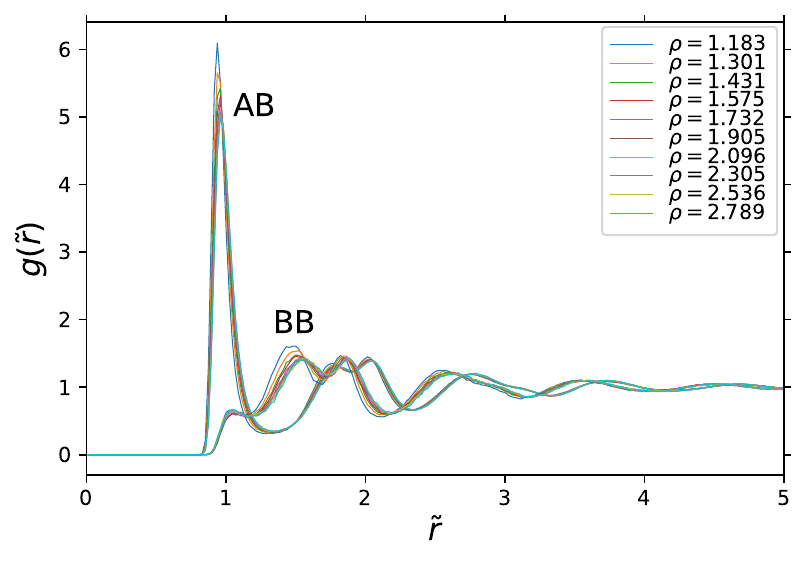}
  \caption{Collapse of $g(\tilde{r})$ for AB and BB pairs as compliment to Fig.~\ref{fig:msd}(b).}
\label{fig:gr_abbb}
\end{figure}

Fig.~\ref{fig:gr_abbb} shows the collapse of the AB and BB radial distribution functions along the isomorph. The collapse here is noticeably worse than for the AA one (Fig.~\ref{fig:msd}(b)); specifically the first peak in the AB case decreases ~15\% as opposed to less than 2\%. Some decrease can be rationalized since the behavior at small $r$ is dominated by the pair potential which becomes effectively softer at higher densities (or, a given value of the reduced-unit $\tilde r$ corresponds to a smaller value of the real distance as density increases, and the effective IPL exponent of the LJ potential decreases with decreasing $r$, approaching 12 in the limit $r\rightarrow 0$). A softer pair potential allows the first peak to extend to shorter distances, and the peak decreases correspondingly to keep the number of neighbors constant.


\begin{thebibliography}{58}
\expandafter\ifx\csname natexlab\endcsname\relax\def\natexlab#1{#1}\fi
\expandafter\ifx\csname bibnamefont\endcsname\relax
  \def\bibnamefont#1{#1}\fi
\expandafter\ifx\csname bibfnamefont\endcsname\relax
  \def\bibfnamefont#1{#1}\fi
\expandafter\ifx\csname citenamefont\endcsname\relax
  \def\citenamefont#1{#1}\fi
\expandafter\ifx\csname url\endcsname\relax
  \def\url#1{\texttt{#1}}\fi
\expandafter\ifx\csname urlprefix\endcsname\relax\def\urlprefix{URL }\fi
\providecommand{\bibinfo}[2]{#2}
\providecommand{\eprint}[2][]{\url{#2}}

\bibitem[{\citenamefont{Gnan et~al.}(2009)\citenamefont{Gnan, Schr{\o}der,
  Pedersen, Bailey, and Dyre}}]{Gnan2009}
\bibinfo{author}{\bibfnamefont{N.}~\bibnamefont{Gnan}},
  \bibinfo{author}{\bibfnamefont{T.~B.} \bibnamefont{Schr{\o}der}},
  \bibinfo{author}{\bibfnamefont{U.~R.} \bibnamefont{Pedersen}},
  \bibinfo{author}{\bibfnamefont{N.~P.} \bibnamefont{Bailey}},
  \bibnamefont{and} \bibinfo{author}{\bibfnamefont{J.~C.} \bibnamefont{Dyre}},
  \bibinfo{journal}{J. Chem. Phys.} \textbf{\bibinfo{volume}{131}},
  \bibinfo{pages}{234504} (\bibinfo{year}{2009}).

\bibitem[{\citenamefont{Bailey et~al.}(2008{\natexlab{a}})\citenamefont{Bailey,
  Pedersen, Gnan, Schr{\o}der, and Dyre}}]{Bailey/others:2008b}
\bibinfo{author}{\bibfnamefont{N.~P.} \bibnamefont{Bailey}},
  \bibinfo{author}{\bibfnamefont{U.~R.} \bibnamefont{Pedersen}},
  \bibinfo{author}{\bibfnamefont{N.}~\bibnamefont{Gnan}},
  \bibinfo{author}{\bibfnamefont{T.~B.} \bibnamefont{Schr{\o}der}},
  \bibnamefont{and} \bibinfo{author}{\bibfnamefont{J.~C.} \bibnamefont{Dyre}},
  \bibinfo{journal}{J. Comp. Phys.} \textbf{\bibinfo{volume}{129}},
  \bibinfo{pages}{184507} (\bibinfo{year}{2008}{\natexlab{a}}).

\bibitem[{\citenamefont{Bailey et~al.}(2008{\natexlab{b}})\citenamefont{Bailey,
  Pedersen, Gnan, Schr{\o}der, and Dyre}}]{Bailey/others:2008c}
\bibinfo{author}{\bibfnamefont{N.~P.} \bibnamefont{Bailey}},
  \bibinfo{author}{\bibfnamefont{U.~R.} \bibnamefont{Pedersen}},
  \bibinfo{author}{\bibfnamefont{N.}~\bibnamefont{Gnan}},
  \bibinfo{author}{\bibfnamefont{T.~B.} \bibnamefont{Schr{\o}der}},
  \bibnamefont{and} \bibinfo{author}{\bibfnamefont{J.~C.} \bibnamefont{Dyre}},
  \bibinfo{journal}{J. Comp. Phys.} \textbf{\bibinfo{volume}{129}},
  \bibinfo{pages}{184508} (\bibinfo{year}{2008}{\natexlab{b}}).

\bibitem[{\citenamefont{Schr{\o}der et~al.}(2009)\citenamefont{Schr{\o}der,
  Bailey, Pedersen, Gnan, and Dyre}}]{Schroder/others:2009b}
\bibinfo{author}{\bibfnamefont{T.~B.} \bibnamefont{Schr{\o}der}},
  \bibinfo{author}{\bibfnamefont{N.~P.} \bibnamefont{Bailey}},
  \bibinfo{author}{\bibfnamefont{U.~R.} \bibnamefont{Pedersen}},
  \bibinfo{author}{\bibfnamefont{N.}~\bibnamefont{Gnan}}, \bibnamefont{and}
  \bibinfo{author}{\bibfnamefont{J.~C.} \bibnamefont{Dyre}},
  \bibinfo{journal}{J. Chem. Phys.} \textbf{\bibinfo{volume}{131}},
  \bibinfo{pages}{234503} (\bibinfo{year}{2009}).

\bibitem[{\citenamefont{Schr{\o}der et~al.}(2011)\citenamefont{Schr{\o}der,
  Gnan, Pedersen, Bailey, and Dyre}}]{Schroder/others:2011}
\bibinfo{author}{\bibfnamefont{T.~B.} \bibnamefont{Schr{\o}der}},
  \bibinfo{author}{\bibfnamefont{N.}~\bibnamefont{Gnan}},
  \bibinfo{author}{\bibfnamefont{U.~R.} \bibnamefont{Pedersen}},
  \bibinfo{author}{\bibfnamefont{N.~P.} \bibnamefont{Bailey}},
  \bibnamefont{and} \bibinfo{author}{\bibfnamefont{J.~C.} \bibnamefont{Dyre}},
  \bibinfo{journal}{J. Chem. Phys.} \textbf{\bibinfo{volume}{134}},
  \bibinfo{pages}{164505} (\bibinfo{year}{2011}).

\bibitem[{\citenamefont{Veldhorst et~al.}(2014)\citenamefont{Veldhorst, Dyre,
  and Schr{\o}der}}]{Veldhorst/Dyre/Schroder:2014}
\bibinfo{author}{\bibfnamefont{A.~A.} \bibnamefont{Veldhorst}},
  \bibinfo{author}{\bibfnamefont{J.~C.} \bibnamefont{Dyre}}, \bibnamefont{and}
  \bibinfo{author}{\bibfnamefont{T.~B.} \bibnamefont{Schr{\o}der}},
  \bibinfo{journal}{J. Chem. Phys.} \textbf{\bibinfo{volume}{141}},
  \bibinfo{pages}{054904} (\bibinfo{year}{2014}).

\bibitem[{\citenamefont{Hummel et~al.}(2015)\citenamefont{Hummel, Kresse, Dyre,
  and Pedersen}}]{Hummel/others:2015}
\bibinfo{author}{\bibfnamefont{F.}~\bibnamefont{Hummel}},
  \bibinfo{author}{\bibfnamefont{G.}~\bibnamefont{Kresse}},
  \bibinfo{author}{\bibfnamefont{J.~C.} \bibnamefont{Dyre}}, \bibnamefont{and}
  \bibinfo{author}{\bibfnamefont{U.~R.} \bibnamefont{Pedersen}},
  \bibinfo{journal}{Phys. Rev. B} \textbf{\bibinfo{volume}{92}},
  \bibinfo{pages}{174116} (\bibinfo{year}{2015}).

\bibitem[{\citenamefont{Gundermann et~al.}(2011)\citenamefont{Gundermann,
  Pedersen, Hecksher, Bailey, Jakobsen, Christensen, Olsen, Schr{\o}der,
  Fragiadakis, Casalini et~al.}}]{Gundermann/others:2011}
\bibinfo{author}{\bibfnamefont{D.}~\bibnamefont{Gundermann}},
  \bibinfo{author}{\bibfnamefont{U.~R.} \bibnamefont{Pedersen}},
  \bibinfo{author}{\bibfnamefont{T.}~\bibnamefont{Hecksher}},
  \bibinfo{author}{\bibfnamefont{N.~P.} \bibnamefont{Bailey}},
  \bibinfo{author}{\bibfnamefont{B.}~\bibnamefont{Jakobsen}},
  \bibinfo{author}{\bibfnamefont{T.}~\bibnamefont{Christensen}},
  \bibinfo{author}{\bibfnamefont{N.~B.} \bibnamefont{Olsen}},
  \bibinfo{author}{\bibfnamefont{T.~B.} \bibnamefont{Schr{\o}der}},
  \bibinfo{author}{\bibfnamefont{D.}~\bibnamefont{Fragiadakis}},
  \bibinfo{author}{\bibfnamefont{R.}~\bibnamefont{Casalini}},
  \bibnamefont{et~al.}, \bibinfo{journal}{Nature Physics}
  \textbf{\bibinfo{volume}{7}}, \bibinfo{pages}{816} (\bibinfo{year}{2011}).

\bibitem[{\citenamefont{Hansen et~al.}(2018)\citenamefont{Hansen, Sanz,
  Adrjanowicz, Frick, and Niss}}]{Hansen2018}
\bibinfo{author}{\bibfnamefont{H.~W.} \bibnamefont{Hansen}},
  \bibinfo{author}{\bibfnamefont{A.}~\bibnamefont{Sanz}},
  \bibinfo{author}{\bibfnamefont{K.}~\bibnamefont{Adrjanowicz}},
  \bibinfo{author}{\bibfnamefont{B.}~\bibnamefont{Frick}}, \bibnamefont{and}
  \bibinfo{author}{\bibfnamefont{K.}~\bibnamefont{Niss}},
  \bibinfo{journal}{Nature Communications} \textbf{\bibinfo{volume}{9}},
  \bibinfo{pages}{518} (\bibinfo{year}{2018}).

\bibitem[{\citenamefont{Pedersen et~al.}(2011)\citenamefont{Pedersen, Gnan,
  Bailey, Schr{\o}der, and Dyre}}]{Pedersen/others:2011}
\bibinfo{author}{\bibfnamefont{U.~R.} \bibnamefont{Pedersen}},
  \bibinfo{author}{\bibfnamefont{N.}~\bibnamefont{Gnan}},
  \bibinfo{author}{\bibfnamefont{N.~P.} \bibnamefont{Bailey}},
  \bibinfo{author}{\bibfnamefont{T.~B.} \bibnamefont{Schr{\o}der}},
  \bibnamefont{and} \bibinfo{author}{\bibfnamefont{J.~C.} \bibnamefont{Dyre}},
  \bibinfo{journal}{J. Non-Cryst Solids} \textbf{\bibinfo{volume}{357}},
  \bibinfo{pages}{320} (\bibinfo{year}{2011}).

\bibitem[{\citenamefont{Dyre}(2014)}]{Dyre:2014}
\bibinfo{author}{\bibfnamefont{J.~C.} \bibnamefont{Dyre}}, \bibinfo{journal}{J.
  Phys. Chem. B} \textbf{\bibinfo{volume}{118}}, \bibinfo{pages}{10007}
  (\bibinfo{year}{2014}).

\bibitem[{\citenamefont{Dyre}(2016)}]{Dyre:2016}
\bibinfo{author}{\bibfnamefont{J.~C.} \bibnamefont{Dyre}}, \bibinfo{journal}{J.
  Phys.: Condens. Mat.} \textbf{\bibinfo{volume}{28}}, \bibinfo{pages}{323001}
  (\bibinfo{year}{2016}).

\bibitem[{\citenamefont{Pedersen et~al.}(2016)\citenamefont{Pedersen,
  Costigliola, Bailey, Schr{\o}der, and Dyre}}]{Pedersen/others:2016a}
\bibinfo{author}{\bibfnamefont{U.~R.} \bibnamefont{Pedersen}},
  \bibinfo{author}{\bibfnamefont{L.}~\bibnamefont{Costigliola}},
  \bibinfo{author}{\bibfnamefont{N.~P.} \bibnamefont{Bailey}},
  \bibinfo{author}{\bibfnamefont{T.~B.} \bibnamefont{Schr{\o}der}},
  \bibnamefont{and} \bibinfo{author}{\bibfnamefont{J.~C.} \bibnamefont{Dyre}},
  \bibinfo{journal}{Nat. Commun.} \textbf{\bibinfo{volume}{7}},
  \bibinfo{pages}{12386} (\bibinfo{year}{2016}).

\bibitem[{\citenamefont{Singh et~al.}(2021)\citenamefont{Singh, Dyre, and
  Pedersen}}]{Singh/Dyre/Pedersen:2021}
\bibinfo{author}{\bibfnamefont{A.~N.} \bibnamefont{Singh}},
  \bibinfo{author}{\bibfnamefont{J.~C.} \bibnamefont{Dyre}}, \bibnamefont{and}
  \bibinfo{author}{\bibfnamefont{U.~R.} \bibnamefont{Pedersen}},
  \bibinfo{journal}{J. Chem. Phys.} \textbf{\bibinfo{volume}{154}},
  \bibinfo{pages}{134501} (\bibinfo{year}{2021}).

\bibitem[{\citenamefont{Dyre}(2020)}]{Dyre2020}
\bibinfo{author}{\bibfnamefont{J.~C.} \bibnamefont{Dyre}}, \bibinfo{journal}{J.
  Chem. Phys.} \textbf{\bibinfo{volume}{153}}, \bibinfo{pages}{134502}
  (\bibinfo{year}{2020}).

\bibitem[{\citenamefont{Separdar et~al.}(2013)\citenamefont{Separdar, Bailey,
  Schr{\o}der, Davatolhagh, and Dyre}}]{Separdar2013}
\bibinfo{author}{\bibfnamefont{L.}~\bibnamefont{Separdar}},
  \bibinfo{author}{\bibfnamefont{N.~P.} \bibnamefont{Bailey}},
  \bibinfo{author}{\bibfnamefont{T.~B.} \bibnamefont{Schr{\o}der}},
  \bibinfo{author}{\bibfnamefont{S.}~\bibnamefont{Davatolhagh}},
  \bibnamefont{and} \bibinfo{author}{\bibfnamefont{J.~C.} \bibnamefont{Dyre}},
  \bibinfo{journal}{J. Chem. Phys.} \textbf{\bibinfo{volume}{138}},
  \bibinfo{pages}{154505} (\bibinfo{year}{2013}).

\bibitem[{\citenamefont{Kob and Andersen}(1994)}]{Kob/Andersen:1994}
\bibinfo{author}{\bibfnamefont{W.}~\bibnamefont{Kob}} \bibnamefont{and}
  \bibinfo{author}{\bibfnamefont{H.~C.} \bibnamefont{Andersen}},
  \bibinfo{journal}{Phys. Rev. Lett.} \textbf{\bibinfo{volume}{73}},
  \bibinfo{pages}{1376} (\bibinfo{year}{1994}).

\bibitem[{\citenamefont{Kob and
  Andersen}(1995{\natexlab{a}})}]{Kob/Andersen:1995a}
\bibinfo{author}{\bibfnamefont{W.}~\bibnamefont{Kob}} \bibnamefont{and}
  \bibinfo{author}{\bibfnamefont{H.~C.} \bibnamefont{Andersen}},
  \bibinfo{journal}{Phys. Rev. E} \textbf{\bibinfo{volume}{51}},
  \bibinfo{pages}{4626} (\bibinfo{year}{1995}{\natexlab{a}}).

\bibitem[{\citenamefont{Kob and
  Andersen}(1995{\natexlab{b}})}]{Kob/Andersen:1995b}
\bibinfo{author}{\bibfnamefont{W.}~\bibnamefont{Kob}} \bibnamefont{and}
  \bibinfo{author}{\bibfnamefont{H.~C.} \bibnamefont{Andersen}},
  \bibinfo{journal}{Phys. Rev. E} \textbf{\bibinfo{volume}{52}},
  \bibinfo{pages}{4134} (\bibinfo{year}{1995}{\natexlab{b}}).

\bibitem[{\citenamefont{Jiang et~al.}(2019)\citenamefont{Jiang, Weeks, and
  Bailey}}]{Jiang2019}
\bibinfo{author}{\bibfnamefont{Y.}~\bibnamefont{Jiang}},
  \bibinfo{author}{\bibfnamefont{E.~R.} \bibnamefont{Weeks}}, \bibnamefont{and}
  \bibinfo{author}{\bibfnamefont{N.~P.} \bibnamefont{Bailey}},
  \bibinfo{journal}{Phys. Rev. E} \textbf{\bibinfo{volume}{100}},
  \bibinfo{pages}{053005} (\bibinfo{year}{2019}).

\bibitem[{\citenamefont{Heyes et~al.}(2019)\citenamefont{Heyes, Dini,
  Costigliola, and Dyre}}]{Heyes/Others:2019}
\bibinfo{author}{\bibfnamefont{D.~M.} \bibnamefont{Heyes}},
  \bibinfo{author}{\bibfnamefont{D.}~\bibnamefont{Dini}},
  \bibinfo{author}{\bibfnamefont{L.}~\bibnamefont{Costigliola}},
  \bibnamefont{and} \bibinfo{author}{\bibfnamefont{J.~C.} \bibnamefont{Dyre}},
  \bibinfo{journal}{J. Chem. Phys.} \textbf{\bibinfo{volume}{151}},
  \bibinfo{pages}{204502} (\bibinfo{year}{2019}).

\bibitem[{\citenamefont{Knudsen et~al.}(2021)\citenamefont{Knudsen, Todd, Dyre,
  and Hansen}}]{Knudsen/Others:2021}
\bibinfo{author}{\bibfnamefont{S.}~\bibnamefont{Knudsen}},
  \bibinfo{author}{\bibfnamefont{B.~D.} \bibnamefont{Todd}},
  \bibinfo{author}{\bibfnamefont{J.~C.} \bibnamefont{Dyre}}, \bibnamefont{and}
  \bibinfo{author}{\bibfnamefont{J.~S.} \bibnamefont{Hansen}},
  \bibinfo{journal}{Phys. Rev. E.} \textbf{\bibinfo{volume}{104}},
  \bibinfo{pages}{054126} (\bibinfo{year}{2021}).

\bibitem[{\citenamefont{Dyre}(2018{\natexlab{a}})}]{Dyre:2018b}
\bibinfo{author}{\bibfnamefont{J.~C.} \bibnamefont{Dyre}}, \bibinfo{journal}{J.
  Chem. Phys.} \textbf{\bibinfo{volume}{149}}, \bibinfo{pages}{210901}
  (\bibinfo{year}{2018}{\natexlab{a}}).

\bibitem[{\citenamefont{Rosenfeld}(1977)}]{Rosenfeld1977}
\bibinfo{author}{\bibfnamefont{Y.}~\bibnamefont{Rosenfeld}},
  \bibinfo{journal}{Phys. Rev. A} \textbf{\bibinfo{volume}{15}},
  \bibinfo{pages}{2545} (\bibinfo{year}{1977}).

\bibitem[{\citenamefont{Gnan et~al.}(2010)\citenamefont{Gnan, Maggi,
  Schr{\o}der, and Dyre}}]{Gnan/Others:2010}
\bibinfo{author}{\bibfnamefont{N.}~\bibnamefont{Gnan}},
  \bibinfo{author}{\bibfnamefont{C.}~\bibnamefont{Maggi}},
  \bibinfo{author}{\bibfnamefont{T.~B.} \bibnamefont{Schr{\o}der}},
  \bibnamefont{and} \bibinfo{author}{\bibfnamefont{J.~C.} \bibnamefont{Dyre}},
  \bibinfo{journal}{Phys. Rev. Lett.} \textbf{\bibinfo{volume}{104}},
  \bibinfo{pages}{125902} (\bibinfo{year}{2010}).

\bibitem[{\citenamefont{Harrison}(1976)}]{Harrison:1976}
\bibinfo{author}{\bibfnamefont{G.}~\bibnamefont{Harrison}},
  \emph{\bibinfo{title}{The dynamic properties of supercooled liquids}}
  (\bibinfo{publisher}{Academic Press}, \bibinfo{year}{1976}).

\bibitem[{\citenamefont{Richert}(2005)}]{Richert:2005}
\bibinfo{author}{\bibfnamefont{R.}~\bibnamefont{Richert}}, \bibinfo{journal}{J.
  Chem. Phys.} \textbf{\bibinfo{volume}{123}}, \bibinfo{pages}{154502}
  (\bibinfo{year}{2005}).

\bibitem[{\citenamefont{Niss and Hecksher}(2018)}]{Niss/Hecksher:2018}
\bibinfo{author}{\bibfnamefont{K.}~\bibnamefont{Niss}} \bibnamefont{and}
  \bibinfo{author}{\bibfnamefont{T.}~\bibnamefont{Hecksher}},
  \bibinfo{journal}{J. Chem. Phys.} \textbf{\bibinfo{volume}{149}},
  \bibinfo{pages}{230901} (\bibinfo{year}{2018}).

\bibitem[{\citenamefont{Toxvaerd and Dyre}(2011)}]{Toxvaerd2011}
\bibinfo{author}{\bibfnamefont{S.}~\bibnamefont{Toxvaerd}} \bibnamefont{and}
  \bibinfo{author}{\bibfnamefont{J.~C.} \bibnamefont{Dyre}},
  \bibinfo{journal}{J. Chem. Phys.} \textbf{\bibinfo{volume}{134}},
  \bibinfo{pages}{081102} (\bibinfo{year}{2011}).

\bibitem[{\citenamefont{Pedersen et~al.}(2018)\citenamefont{Pedersen,
  Schr{\o}der, and Dyre}}]{Pedersen2018}
\bibinfo{author}{\bibfnamefont{U.~R.} \bibnamefont{Pedersen}},
  \bibinfo{author}{\bibfnamefont{T.~B.} \bibnamefont{Schr{\o}der}},
  \bibnamefont{and} \bibinfo{author}{\bibfnamefont{J.~C.} \bibnamefont{Dyre}},
  \bibinfo{journal}{Phys. Rev. Lett.} \textbf{\bibinfo{volume}{120}},
  \bibinfo{pages}{165501} (\bibinfo{year}{2018}).

\bibitem[{\citenamefont{Evans and Morriss}(1984)}]{Evans/Morriss:1984}
\bibinfo{author}{\bibfnamefont{D.~J.} \bibnamefont{Evans}} \bibnamefont{and}
  \bibinfo{author}{\bibfnamefont{G.~P.} \bibnamefont{Morriss}},
  \bibinfo{journal}{Phys.\ Rev.\ A} \textbf{\bibinfo{volume}{30}},
  \bibinfo{pages}{1528} (\bibinfo{year}{1984}).

\bibitem[{\citenamefont{Ladd}(1984)}]{Ladd:1984}
\bibinfo{author}{\bibfnamefont{A.~J.~C.} \bibnamefont{Ladd}},
  \bibinfo{journal}{Mol. Phys.} \textbf{\bibinfo{volume}{53}},
  \bibinfo{pages}{459} (\bibinfo{year}{1984}).

\bibitem[{\citenamefont{Lees and Edwards}(1972)}]{Lees/Edwards:1972}
\bibinfo{author}{\bibfnamefont{A.~W.} \bibnamefont{Lees}} \bibnamefont{and}
  \bibinfo{author}{\bibfnamefont{S.~F.} \bibnamefont{Edwards}},
  \bibinfo{journal}{J. Phys. C: Solid State Phys} \textbf{\bibinfo{volume}{5}},
  \bibinfo{pages}{1921} (\bibinfo{year}{1972}).

\bibitem[{\citenamefont{Allen and Tildesley}(1987)}]{Allen/Tildesley:1987}
\bibinfo{author}{\bibfnamefont{M.~P.} \bibnamefont{Allen}} \bibnamefont{and}
  \bibinfo{author}{\bibfnamefont{D.~J.} \bibnamefont{Tildesley}},
  \emph{\bibinfo{title}{Computer Simulation of Liquids}}
  (\bibinfo{publisher}{Oxford University Press}, \bibinfo{year}{1987}).

\bibitem[{\citenamefont{Evans and Morriss}(1986)}]{Evans/Morriss:1986}
\bibinfo{author}{\bibfnamefont{D.~J.} \bibnamefont{Evans}} \bibnamefont{and}
  \bibinfo{author}{\bibfnamefont{G.~P.} \bibnamefont{Morriss}},
  \bibinfo{journal}{Phys. Rev. Lett.} \textbf{\bibinfo{volume}{56}},
  \bibinfo{pages}{2172} (\bibinfo{year}{1986}).

\bibitem[{\citenamefont{Baranyai}(2000)}]{Baranyai:2000}
\bibinfo{author}{\bibfnamefont{A.}~\bibnamefont{Baranyai}},
  \bibinfo{journal}{Phys. Rev. E Stat. Phys.} \textbf{\bibinfo{volume}{62}},
  \bibinfo{pages}{5989} (\bibinfo{year}{2000}).

\bibitem[{\citenamefont{Hoover et~al.}(2008)\citenamefont{Hoover, Hoover, and
  Petravic}}]{Hoover/Hoover/Petravic:2008}
\bibinfo{author}{\bibfnamefont{W.~G.} \bibnamefont{Hoover}},
  \bibinfo{author}{\bibfnamefont{C.~G.} \bibnamefont{Hoover}},
  \bibnamefont{and} \bibinfo{author}{\bibfnamefont{J.}~\bibnamefont{Petravic}},
  \bibinfo{journal}{Phys. Rev. E Stat. Nonlin. Soft Matter Phys.}
  \textbf{\bibinfo{volume}{78}}, \bibinfo{pages}{046701}
  (\bibinfo{year}{2008}).

\bibitem[{\citenamefont{Morriss and Dettmann}(1998)}]{Morriss/Dettmann:1998}
\bibinfo{author}{\bibfnamefont{G.~P.} \bibnamefont{Morriss}} \bibnamefont{and}
  \bibinfo{author}{\bibfnamefont{C.~P.} \bibnamefont{Dettmann}},
  \bibinfo{journal}{Chaos} \textbf{\bibinfo{volume}{8}}, \bibinfo{pages}{321}
  (\bibinfo{year}{1998}).

\bibitem[{\citenamefont{Singh et~al.}(2020)\citenamefont{Singh, Ozawa, and
  Berthier}}]{Singh2020}
\bibinfo{author}{\bibfnamefont{M.}~\bibnamefont{Singh}},
  \bibinfo{author}{\bibfnamefont{M.}~\bibnamefont{Ozawa}}, \bibnamefont{and}
  \bibinfo{author}{\bibfnamefont{L.}~\bibnamefont{Berthier}},
  \bibinfo{journal}{Phys. Rev. Mater.} \textbf{\bibinfo{volume}{4}},
  \bibinfo{pages}{025603} (\bibinfo{year}{2020}).

\bibitem[{\citenamefont{Taylor}(1996)}]{Taylor:1996}
\bibinfo{author}{\bibfnamefont{J.}~\bibnamefont{Taylor}},
  \emph{\bibinfo{title}{{An Introduction to Error Analysis: The Study of
  Uncertainties in Physical Measurements}}} (\bibinfo{publisher}{University
  Science Books}, \bibinfo{year}{1996}), \bibinfo{edition}{2nd} ed.

\bibitem[{\citenamefont{Langer and Liu}(1997)}]{Langer/Liu:1997}
\bibinfo{author}{\bibfnamefont{S.~A.} \bibnamefont{Langer}} \bibnamefont{and}
  \bibinfo{author}{\bibfnamefont{A.~J.} \bibnamefont{Liu}},
  \bibinfo{journal}{J. Phys. Chem. B} \textbf{\bibinfo{volume}{101}},
  \bibinfo{pages}{8667–8671} (\bibinfo{year}{1997}).

\bibitem[{\citenamefont{Goldenberg et~al.}(2007)\citenamefont{Goldenberg,
  Tanguy, and Barrat}}]{Goldenberg/Tanguy/Barrat:2007}
\bibinfo{author}{\bibfnamefont{C.}~\bibnamefont{Goldenberg}},
  \bibinfo{author}{\bibfnamefont{A.}~\bibnamefont{Tanguy}}, \bibnamefont{and}
  \bibinfo{author}{\bibfnamefont{J.-L.} \bibnamefont{Barrat}},
  \bibinfo{journal}{Europhys. Lett.} \textbf{\bibinfo{volume}{80}},
  \bibinfo{pages}{16003} (\bibinfo{year}{2007}).

\bibitem[{\citenamefont{Chen et~al.}(2010)\citenamefont{Chen, Semwogerere,
  Sato, Breedveld, and Weeks}}]{Chen2010}
\bibinfo{author}{\bibfnamefont{D.}~\bibnamefont{Chen}},
  \bibinfo{author}{\bibfnamefont{D.}~\bibnamefont{Semwogerere}},
  \bibinfo{author}{\bibfnamefont{J.}~\bibnamefont{Sato}},
  \bibinfo{author}{\bibfnamefont{V.}~\bibnamefont{Breedveld}},
  \bibnamefont{and} \bibinfo{author}{\bibfnamefont{E.~R.} \bibnamefont{Weeks}},
  \bibinfo{journal}{Phys. Rev. E} \textbf{\bibinfo{volume}{81}},
  \bibinfo{pages}{011403} (\bibinfo{year}{2010}).

\bibitem[{\citenamefont{Bailey et~al.}(2006)\citenamefont{Bailey, Schi{\o}tz,
  and Jacobsen}}]{Bailey/Schiotz/Jacobsen:2006}
\bibinfo{author}{\bibfnamefont{N.~P.} \bibnamefont{Bailey}},
  \bibinfo{author}{\bibfnamefont{J.}~\bibnamefont{Schi{\o}tz}},
  \bibnamefont{and} \bibinfo{author}{\bibfnamefont{K.~W.}
  \bibnamefont{Jacobsen}}, \bibinfo{journal}{Phys. Rev. B}
  \textbf{\bibinfo{volume}{73}}, \bibinfo{pages}{064108}
  (\bibinfo{year}{2006}).

\bibitem[{\citenamefont{Ovarlez et~al.}(2009)\citenamefont{Ovarlez, Rodts,
  Chateau, and Coussot}}]{Ovarlez/Others:2009}
\bibinfo{author}{\bibfnamefont{G.}~\bibnamefont{Ovarlez}},
  \bibinfo{author}{\bibfnamefont{S.}~\bibnamefont{Rodts}},
  \bibinfo{author}{\bibfnamefont{X.}~\bibnamefont{Chateau}}, \bibnamefont{and}
  \bibinfo{author}{\bibfnamefont{P.}~\bibnamefont{Coussot}},
  \bibinfo{journal}{Rheol. Acta} \textbf{\bibinfo{volume}{48}},
  \bibinfo{pages}{831} (\bibinfo{year}{2009}).

\bibitem[{\citenamefont{Martin and Thomas~Hu}(2012)}]{Martin/Hu:2012}
\bibinfo{author}{\bibfnamefont{J.~D.} \bibnamefont{Martin}} \bibnamefont{and}
  \bibinfo{author}{\bibfnamefont{Y.}~\bibnamefont{Thomas~Hu}},
  \bibinfo{journal}{Soft matter} \textbf{\bibinfo{volume}{8}},
  \bibinfo{pages}{6940} (\bibinfo{year}{2012}).

\bibitem[{\citenamefont{Greer et~al.}(2013)\citenamefont{Greer, Cheng, and
  Ma}}]{Greer/Cheng/Ma:2013}
\bibinfo{author}{\bibfnamefont{A.~L.} \bibnamefont{Greer}},
  \bibinfo{author}{\bibfnamefont{Y.~Q.} \bibnamefont{Cheng}}, \bibnamefont{and}
  \bibinfo{author}{\bibfnamefont{E.}~\bibnamefont{Ma}},
  \bibinfo{journal}{Materials Science and Engineering: R: Reports}
  \textbf{\bibinfo{volume}{74}}, \bibinfo{pages}{71} (\bibinfo{year}{2013}).

\bibitem[{\citenamefont{Bacher et~al.}(2018)\citenamefont{Bacher, Schr{\o}der,
  and Dyre}}]{Bacher/Schroder/Dyre:2018}
\bibinfo{author}{\bibfnamefont{A.~K.} \bibnamefont{Bacher}},
  \bibinfo{author}{\bibfnamefont{T.~B.} \bibnamefont{Schr{\o}der}},
  \bibnamefont{and} \bibinfo{author}{\bibfnamefont{J.~C.} \bibnamefont{Dyre}},
  \bibinfo{journal}{J. Chem. Phys.} \textbf{\bibinfo{volume}{149}},
  \bibinfo{pages}{114502} (\bibinfo{year}{2018}).

\bibitem[{\citenamefont{Friedeheim et~al.}(2019)\citenamefont{Friedeheim, Dyre,
  and Bailey}}]{Friedeheim2019}
\bibinfo{author}{\bibfnamefont{L.}~\bibnamefont{Friedeheim}},
  \bibinfo{author}{\bibfnamefont{J.~C.} \bibnamefont{Dyre}}, \bibnamefont{and}
  \bibinfo{author}{\bibfnamefont{N.~P.} \bibnamefont{Bailey}},
  \bibinfo{journal}{Phys. Rev. E} \textbf{\bibinfo{volume}{99}},
  \bibinfo{pages}{022142} (\bibinfo{year}{2019}).

\bibitem[{\citenamefont{Bagchi et~al.}(1996)\citenamefont{Bagchi,
  Balasubramanian, Mundy, and Klein}}]{Bagchi/Others:1996}
\bibinfo{author}{\bibfnamefont{K.}~\bibnamefont{Bagchi}},
  \bibinfo{author}{\bibfnamefont{S.}~\bibnamefont{Balasubramanian}},
  \bibinfo{author}{\bibfnamefont{C.~J.} \bibnamefont{Mundy}}, \bibnamefont{and}
  \bibinfo{author}{\bibfnamefont{M.~L.} \bibnamefont{Klein}},
  \bibinfo{journal}{J. Chem. Phys.} \textbf{\bibinfo{volume}{105}},
  \bibinfo{pages}{11183} (\bibinfo{year}{1996}).

\bibitem[{\citenamefont{Voigtmann}(2014)}]{Voigtmann:2014}
\bibinfo{author}{\bibfnamefont{T.}~\bibnamefont{Voigtmann}},
  \bibinfo{journal}{Curr. Opin. Colloid Interface Sci.}
  \textbf{\bibinfo{volume}{19}}, \bibinfo{pages}{549} (\bibinfo{year}{2014}).

\bibitem[{\citenamefont{Dyre}(2018{\natexlab{b}})}]{Dyre:2018a}
\bibinfo{author}{\bibfnamefont{J.~C.} \bibnamefont{Dyre}}, \bibinfo{journal}{J.
  Chem. Phys.} \textbf{\bibinfo{volume}{148}}, \bibinfo{pages}{154502}
  (\bibinfo{year}{2018}{\natexlab{b}}).

\bibitem[{\citenamefont{Schr{\o}der}(2022)}]{Schroder:2022}
\bibinfo{author}{\bibfnamefont{T.~B.} \bibnamefont{Schr{\o}der}},
  \bibinfo{journal}{Phys. Rev. Lett.} \textbf{\bibinfo{volume}{129}},
  \bibinfo{pages}{245501} (\bibinfo{year}{2022}).

\bibitem[{\citenamefont{Attia et~al.}(2021)\citenamefont{Attia, Dyre, and
  Pedersen}}]{Attia/Dyre/Pedersen:2021}
\bibinfo{author}{\bibfnamefont{E.}~\bibnamefont{Attia}},
  \bibinfo{author}{\bibfnamefont{J.~C.} \bibnamefont{Dyre}}, \bibnamefont{and}
  \bibinfo{author}{\bibfnamefont{U.~R.} \bibnamefont{Pedersen}},
  \bibinfo{journal}{Phys. Rev. E} \textbf{\bibinfo{volume}{103}},
  \bibinfo{pages}{062140} (\bibinfo{year}{2021}).

\bibitem[{\citenamefont{Schr{\o}der and Dyre}(2014)}]{Schroder2014}
\bibinfo{author}{\bibfnamefont{T.~B.} \bibnamefont{Schr{\o}der}}
  \bibnamefont{and} \bibinfo{author}{\bibfnamefont{J.~C.} \bibnamefont{Dyre}},
  \bibinfo{journal}{J. Chem. Phys.} \textbf{\bibinfo{volume}{141}},
  \bibinfo{pages}{204502} (\bibinfo{year}{2014}).

\bibitem[{\citenamefont{Ingebrigtsen et~al.}(2012)\citenamefont{Ingebrigtsen,
  B{\o}hling, Schr{\o}der, and Dyre}}]{Ingebrigtsen2012}
\bibinfo{author}{\bibfnamefont{T.~S.} \bibnamefont{Ingebrigtsen}},
  \bibinfo{author}{\bibfnamefont{L.}~\bibnamefont{B{\o}hling}},
  \bibinfo{author}{\bibfnamefont{T.~B.} \bibnamefont{Schr{\o}der}},
  \bibnamefont{and} \bibinfo{author}{\bibfnamefont{J.~C.} \bibnamefont{Dyre}},
  \bibinfo{journal}{J. Chem. Phys.} \textbf{\bibinfo{volume}{136}},
  \bibinfo{pages}{061102} (\bibinfo{year}{2012}).

\bibitem[{\citenamefont{B{\o}hling et~al.}(2014)\citenamefont{B{\o}hling,
  Bailey, Schr{\o}der, and Dyre}}]{Bohling2014}
\bibinfo{author}{\bibfnamefont{L.}~\bibnamefont{B{\o}hling}},
  \bibinfo{author}{\bibfnamefont{N.~P.} \bibnamefont{Bailey}},
  \bibinfo{author}{\bibfnamefont{T.~B.} \bibnamefont{Schr{\o}der}},
  \bibnamefont{and} \bibinfo{author}{\bibfnamefont{J.~C.} \bibnamefont{Dyre}},
  \bibinfo{journal}{J. Chem. Phys.} \textbf{\bibinfo{volume}{140}},
  \bibinfo{pages}{124510} (\bibinfo{year}{2014}).

\bibitem[{\citenamefont{B{\o}hling et~al.}(2012)\citenamefont{B{\o}hling,
  Ingebrigtsen, Grzybowski, Paluch, Dyre, and Schr{\o}der}}]{Bohling2012}
\bibinfo{author}{\bibfnamefont{L.}~\bibnamefont{B{\o}hling}},
  \bibinfo{author}{\bibfnamefont{T.~S.} \bibnamefont{Ingebrigtsen}},
  \bibinfo{author}{\bibfnamefont{A.}~\bibnamefont{Grzybowski}},
  \bibinfo{author}{\bibfnamefont{M.}~\bibnamefont{Paluch}},
  \bibinfo{author}{\bibfnamefont{J.~C.} \bibnamefont{Dyre}}, \bibnamefont{and}
  \bibinfo{author}{\bibfnamefont{T.~B.} \bibnamefont{Schr{\o}der}},
  \bibinfo{journal}{New J. Phys.} \textbf{\bibinfo{volume}{14}},
  \bibinfo{pages}{113035} (\bibinfo{year}{2012}).

\end{thebibliography}
\end{document}